\documentclass[aps,pra,twocolumn,reprint,floatfix, superscriptaddress]{revtex4-2}

\usepackage{subfigure}
\usepackage{psfrag,graphicx}
\usepackage{pict2e}
\usepackage{dcolumn}
\usepackage{amsmath,amssymb,braket}
\usepackage{bm}
\usepackage{color}
\usepackage{latexsym}
\usepackage{epstopdf}
\usepackage{color}
\usepackage[english]{babel}
\UseRawInputEncoding
\usepackage{amsfonts}
\usepackage{bm}
\usepackage{natbib}
\usepackage{bbold}

\usepackage{enumerate}
\definecolor{myblue}{rgb}{0.2,0.2,0.8}
\definecolor{myzard}{cmyk}{0,0,0.05,0}
\definecolor{mywhite}{rgb}{1,1,1}
\definecolor{mywhite}{rgb}{1,1,1}
\definecolor{myred}{rgb}{1,0.,0.3}
\usepackage[colorlinks=true,citecolor=myblue,linkcolor=myred]{hyperref}

\definecolor{darkgreen}{rgb}{0.0, 0.4, 0.26}

\definecolor{mygrey}{gray}{0.35}
\definecolor{myblue}{rgb}{0.2,0.2,0.8}
\definecolor{myzard}{cmyk}{0,0,0.05,0}
\definecolor{mywhite}{rgb}{1,1,1}
\definecolor{mywhite}{rgb}{1,1,1}
\definecolor{myred}{rgb}{1,0.,0.3}

\def\be{\begin{equation}}
\def\ee{\end{equation}}
\def\ba{\begin{align}}
\def\enda{\end{align}}
\def\bi{\begin{itemize}}
\def\ei{\end{itemize}}

\def\beq{\begin{equation}}
\def\beq{\begin{equation}}
\def\eeq{\end{equation}}

\newcommand\numberthis{\addtocounter{equation}{1}\tag{\theequation}}

\begin{document}

\title{Qubit-photon bound states in topological waveguides with long-range hoppings}

\author{C.~Vega}
\email{carlos.vega@iff.csic.es}
\affiliation{Institute of Fundamental Physics IFF-CSIC, Calle Serrano 113b, 28006 Madrid, Spain.}

\author{M.~Bello}
\email{miguel.bello@mpq.mpg.de}
\affiliation{Max-Planck-Institut f\"ur Quantenoptik, Hans-Kopfermann-Strasse 1, 85748 Garching, Germany.}

\author{D.~Porras}
\email{diego.porras@csic.es}
\affiliation{Institute of Fundamental Physics IFF-CSIC, Calle Serrano 113b, 28006 Madrid, Spain.}

\author{A.~González-Tudela}
\email{a.gonzalez.tudela@csic.es}
\affiliation{Institute of Fundamental Physics IFF-CSIC, Calle Serrano 113b, 28006 Madrid, Spain.}

\begin{abstract}
Quantum emitters interacting with photonic band-gap materials lead to the appearance of qubit-photon bound states that mediate decoherence-free, tunable emitter-emitter interactions. Recently, it has been shown that when these band-gaps have a topological origin, like in the photonic SSH model, these qubit-photon bound states feature chiral shapes and certain robustness to disorder. In this work, we consider a more general situation where the emitters interact with an extended SSH photonic model with longer range hoppings that displays a richer phase diagram than its nearest-neighbour counterpart, e.g., phases with larger winding numbers. In particular, we first study the features of the qubit-photon bound states when the emitters couple to the bulk modes in the different phases, discern its connection with the topological invariant, and show how to further tune their shape through the use of giant atoms, i.e., non-local couplings. Then, we consider the coupling of emitters to the edge modes appearing in the different topological phases. Here, we show that giant-atom dynamics can distinguish between all different topological phases, in contrast to the case with local couplings. Finally, we provide a possible experimental implementation of the model based on periodic modulations of circuit QED systems. Our work enriches the understanding of the interplay between topological photonics and quantum optics.
\end{abstract}

\maketitle

\section{Introduction}

The possibility to engineer photonic band-gaps with periodically patterned materials~\cite{Yablonovitch1991} enabled an unprecedented control of the flow of flight in classical photonics~\cite{Joannopoulos2011}. In quantum optics, such band-gaps not only increase the lifetime of emitters through the modification of the local density of states~\cite{purcell46a}, but can even lead to an incomplete spontaneous emission dynamics known as fractional decay~\cite{john94a}. The latter occurs when the emitter's transition frequency lies deep in the band-gap, because then the photon that should relax into the bath has an energy that is not allowed to propagate. In those situations, the photon becomes localized around the emitter~\cite{bykov75a,john90a,kurizki90a} forming what has been labeled as qubit-photon bound states~\cite{douglas15a,Gonzalez-Tudela2015b,Shi2016,calajo16a,Shi2018,liu17a,Sundaresan2019,Mirhosseini2018a,sanchezburillo17a,Roman-Roche2020}. Beyond their fundamental interest, these bound states have recently attracted a lot of attention because they can mediate coherent and tunable interactions between emitters, which can be harnessed for simulating frustrated quantum magnetism problems~\cite{douglas15a,Gonzalez-Tudela2015b}. This has triggered many experiments to observe them, not only in standard photonic crystals~\cite{hood16a}, but also in other platforms that mimic band-gap physics, such as circuit QED~\cite{liu17a,Sundaresan2019,Mirhosseini2018a} or state-dependent optical lattices~\cite{krinner18a}.

The recently discovered topological photonic insulators (see Refs.~\cite{ozawa19a,Rider2019,Smirnova2020} for updated reviews) are one of these systems where photonic band-gaps appear. Their particularity is that the forbidden-energy region appears between two bands that can be characterized by an integer number called the topological invariant, e.g., winding number $\mathcal{W}$~\cite{Ryu2002}, which determines how many topological edge states will emerge in finite systems. A natural question that scientists have begun studying is what quantum optical phenomena occur when coupling quantum emitters to such systems~\cite{barik18a,Barik2020,Mehrabad2019a,Bello2019,Kim2020b,Leonforte2020b,DeBernardis2021,Garcia-Elcano2020,Garcia-Elcano2021} and, in particular, its impact on the emergent qubit-photon bound states. For example, already for the simplest instance of one-dimensional topological photonic insulator, i.e., the SSH model~\cite{ryu10a} which has a two phases with $\mathcal{W}=1,0$, it was predicted~\cite{Bello2019} and later experimentally confirmed~\cite{Kim2020b}, that the qubit-photon bound states display a chiral shape and inherit certain robustness to disorder from the bath. The underlying reason of these remarkable features is that the bound-state builds up from the topological edge modes of the ``broken" photonic lattice that appears if one introduces a vacancy-defect at the position of the emitter~\cite{Leonforte2020b}. Since these edge modes are strongly linked to the value of the topological invariant of the system, an intriguing question to be explored is to consider the case of models with $\mathcal{W}>1$, and see how these qubit-photon bound states change.

In this work, we study this question by considering a photonic lattice with dimerized next-to-next-to-nearest neighbour hoppings, which displays phases with $\mathcal{W} = 0, \pm1, 2$ depending on the parameters of the model~\cite{Li2014,Li2018,Maffei2018,Perez-Gonzalez2019}. We show how topologically robust qubit-photon bound states emerge along all the phases of the diagram, displaying different spatial shapes depending of the parameters of the model. We further show how these
shapes can be controlled using non-local light-matter
couplings~\cite{kockum18a,FriskKockum2021,Kannan2020,Wang2021a}. We also study finite-system effects and
illustrate the emitter’s dynamics in the different phases
when it couples to the edge of the photonic chain. Finally,
we also discuss a possible experimental implementation
based on circuit QED platforms.
The manuscript is structured as follows: in Section~\ref{sec:system} we explain the system under study. We first analyze in detail the photonic lattice Hamiltonian, discussing its band-structure, phase diagram, and edge states. Then, we consider the full light-matter interaction Hamiltonian, and study the emitter expected lifetimes and Lamb-shifts coming from the interaction with the bath. In Section~\ref{sec:bulk}, we study the properties of the emergent qubit-photon bound states when the emitters are coupled to the bulk of the system, with their energies lying in the band-gap. Then, in Section~\ref{sec:edge} we study the case where the emitters couple to the edge of the chain and see how the presence of a different number of edge-states modifies their dynamics. Finally, in Section~\ref{sec:exp} we discuss a possible experimental realization, and summarize our findings in Section~\ref{sec:conclu}.

\section{Light-matter interactions in extended-SSH models~\label{sec:system}}

In this section, we present and analyze the features of the photonic environment considered in this work, in ~\ref{sec:bath}. Afterwards, in~\ref{sec:coupling_emitters}, we introduce the coupling of a collection of quantum emitters (emitters) to the photonic environment described by this model, and analyse the
renormalization effects that the bath induces in the energy and lifetime of a single emitter.

\subsection{Extended-SSH models: band-structure, phase-diagram, and finite-size effects~\label{sec:bath}}

\begin{figure}[tb]
\includegraphics[width=\columnwidth]{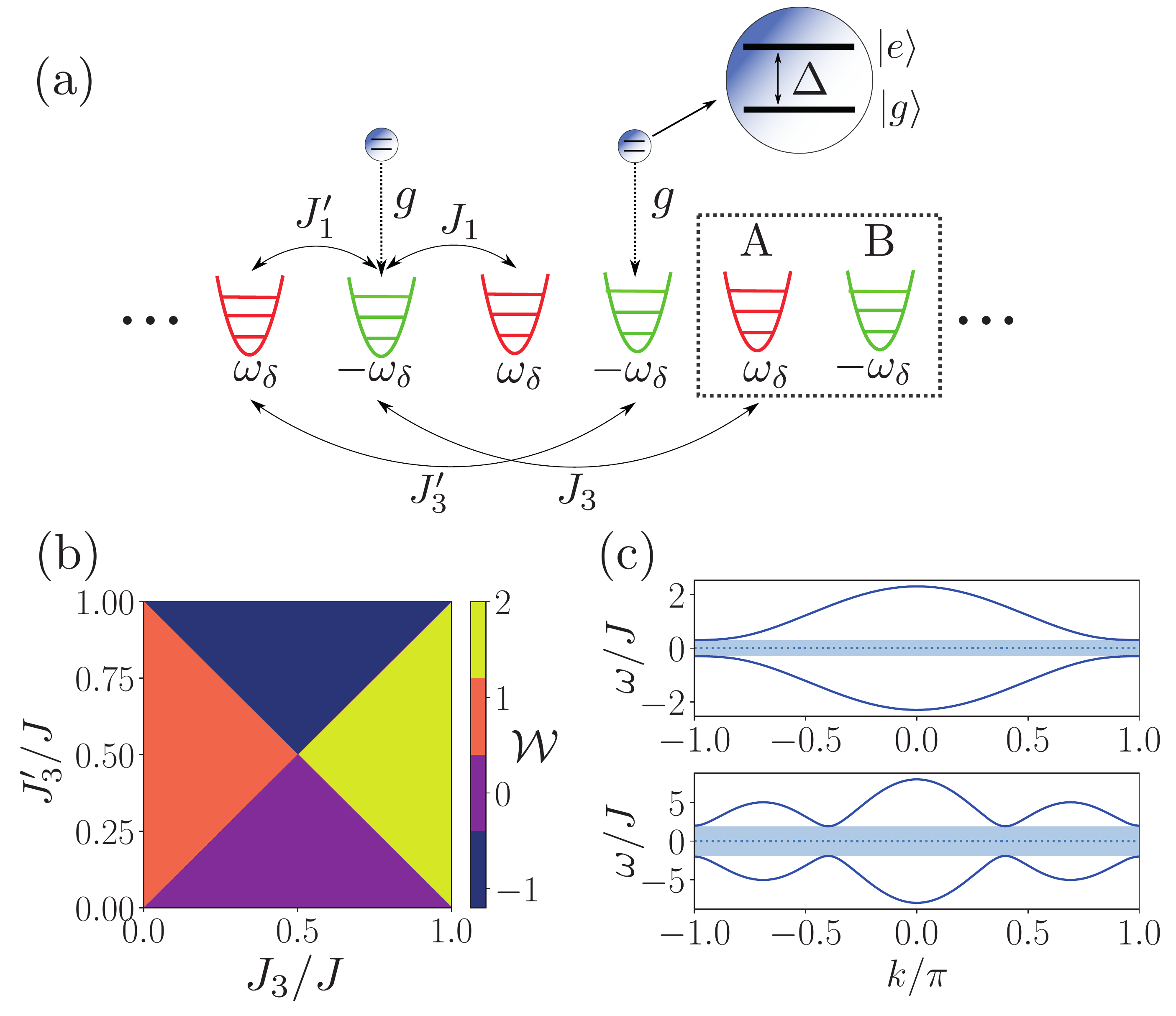}
\caption{a) Scheme of the light-matter setup: The photonic bath is composed as a series of coupled resonators, depicted in green and red, corresponding to crystal sublattices A and B respectively. The energy of the resonators of the two sublattices are given by $\omega_c\pm\omega_\delta$, respectively. The unit cell of the array is represented within a dotted box. Arrows between sites depict hopping amplitudes in the tight-binding Hamiltonian. Two quantum emitters coupled to the lattice with coupling constant $g$ are represented, as well as the emitter level structure. b) Winding number $\mathcal{W}$ of the different phases of the extended-SSH model in the $J_1=J_1^\prime\equiv J$ regime. Parameters $J_3$ and $J_3^\prime$ are expressed in units of $J$. c) Band structure in two scenarios, when either long-range hoppings are small $(J_3, J_3^\prime) = (0, 0.3)$ (up), and when they are large $(J_3, J_3^\prime)=(4, 2)$ (down). In the first case, the band structure does not qualitatively differ from the SSH one, but in the latter, local maxima appear in the bands.}
\label{phase_diagram}
\end{figure}

Along this manuscript, we will use a discrete photonic lattice model description for the photonic bath as depicted in Fig.~\ref{phase_diagram}(a). It is composed by $N$ unit cells, each containing two sites which correspond to bosonic (creation) annhilation operators $a_j^{(\dagger)},b_j^{(\dagger)}$ where $j$ is an index indicating which unit-cell the modes belong to. The Hamiltonian associated to this bipartite photonic lattice can be written as follows:
\begin{align*}
\mathcal{H}_\text{bath} =&\quad \sum_{j=1}^{N} \left\lbrace\omega_c\left(a_j^\dagger a_j + b_j^\dagger b_j\right) 
+ \omega_\delta\left( a_j^\dagger a_j - b_j^\dagger b_j\right)\right\rbrace\\
&+\sum_{j=1}^{N}\Big\{ J_1^\prime a_j^\dagger b_j + J_1 b_j^\dagger a_{j+1}\\
&+J_3^\prime a_j^\dagger b_{j+1} + J_3 b_j^\dagger a_{j+2}+\text{H.c.}\Big\}\numberthis\,.
\end{align*}
Here, $\omega_c$ is the overall energy reference of the problem, we will set it to zero and refer all the other energies with respect to it; $\omega_\delta$ is an staggered-energy offset between the A/B sublattice modes; and $J_{1} (J_1^\prime)$ [$J_{3} (J_3^\prime)$] are the [third] first neighbour hoppings between the B-A (A-B) sublattices, respectively. Using periodic boundary conditions, one can define the photonic operators in momentum space as follows:
\begin{equation}
a_k^\dagger=\frac{1}{\sqrt{N}}\sum_{j=1}^{N}e^{ikj}a_j^\dagger\;\text{ and }\;b_k^\dagger=\frac{1}{\sqrt{N}}\sum_{j=1}^{N}e^{ikj}b_j^\dagger\;,
\end{equation}
and rewrite the bath Hamiltonian as follows:
\begin{align}
\mathcal{H}_\text{bath} =\sum_k \;
   \begin{pmatrix}
   a_k^\dagger & b_k^\dagger
   \end{pmatrix}\;
   H_\text{bath}(k)
   \begin{pmatrix}
   a_k \\ 
   b_k
   \end{pmatrix}\;,   
\end{align}
where $H_\text{bath}(k)$ is a $2\times 2$ matrix that can be expanded in terms of Pauli matrices $H_\text{bath}(k)=\sum_{\alpha=x,y,z}d_\alpha(k)\sigma_\alpha$ as follows:
\begin{align}
    d_x(k)&=J_1^\prime+J_1\cos(k)+J_3^\prime \cos(k)+J_3\cos(2k)\,,\\
    d_y(k)&=J_1\sin(k)-J_3^\prime\sin(k)+J_3\sin(2k)\,,\\
    d_z(k)&=\omega_\delta\,.
\end{align}

Using these expressions, the bath can be readily diagonalized as $H_\text{bath}=\sum_k\left(\omega_u(k)u_k^\dagger u_k+\omega_l(k)l_k^\dagger l_k\right)$, leading to two bands with energy dispersion $\omega_{u/d}(k)=\pm\sqrt{\sum_\alpha d_\alpha^2(k)}=\pm \omega(k)$. The choice of this very general model allows us to capture very different situations, including topologically non-trivial regimes, that we will study along the manuscript:
\begin{itemize}

    \item When $\omega_\delta=J_3^\prime=J_3=0$ and $J_1\neq J^\prime_1\neq0$, one obtains the photonic analogue of the SSH model~\cite{Su1979, ryu00a}, where two separate energy bands appear around the bath energy reference. One of the main properties of this model is that thanks to its chiral (sublattice) symmetry ($d_z(k)\equiv 0$), one can define a topological invariant, i.e., the winding number $\mathcal{W}$, that counts how many times the vector $\Vec{d}(k)=\left(d_x(k), d_y(k)\right)$ winds around the origin as $k$ swipes the Brillouin zone, and which is related to the number of topologically-protected edge states that appear in finite systems with open boundary conditions~\cite{Ryu2002}. This model belongs to the BDI class of the topological classification of phases~\cite{ryu00a}, featuring both topologically trivial and non-trivial phases with $\mathcal{W}=0,1$, respectively, depending on the relative value of $J_1,J_1^\prime$.
    
    \item An interesting variation of the previous case can be done by adding staggered third-neighbour hoppings, $J_3\neq J_3^\prime\neq0$. This model is referred to in the literature as the extended SSH model~\cite{Li2014,Li2018,Maffei2018,Perez-Gonzalez2019}, a notation that we will also use along the whole manuscript. These additional hoppings are chosen so that they still preserve the chiral symmetry of the model. However, they allow for phases with larger values of the winding number $\mathcal{W}= 2, \pm 1, 0$. Note, that adding second-neighbour hoppings (in fact, any even hopping) breaks the chiral symmetry of the model, and change its Atland-Zirnbauer topological class~\cite{ryu10a} from BDI to AI, which is trivial in the case of one-dimensional systems.
    
    \item Another way of making the system topologically trivial without adding any longer range or staggered hoppings ($J_1=J_1^\prime$ and $J_3=J^\prime_3=0$) is by considering staggered-energies, $\omega_\delta\neq 0$. This model also displays a symmetric two-band spectrum around the bath energy reference, like in the previous topological cases, but its band-gap has now a completely (topologically) trivial origin. For this reason, we will use this model in section~\ref{sec:robustness} to unravel the role of topological/bipartite nature of the bath in the robustness of the emergent qubit-photon bound states.
    
\end{itemize}

\begin{figure}[tb]
\includegraphics[width=0.99\columnwidth]{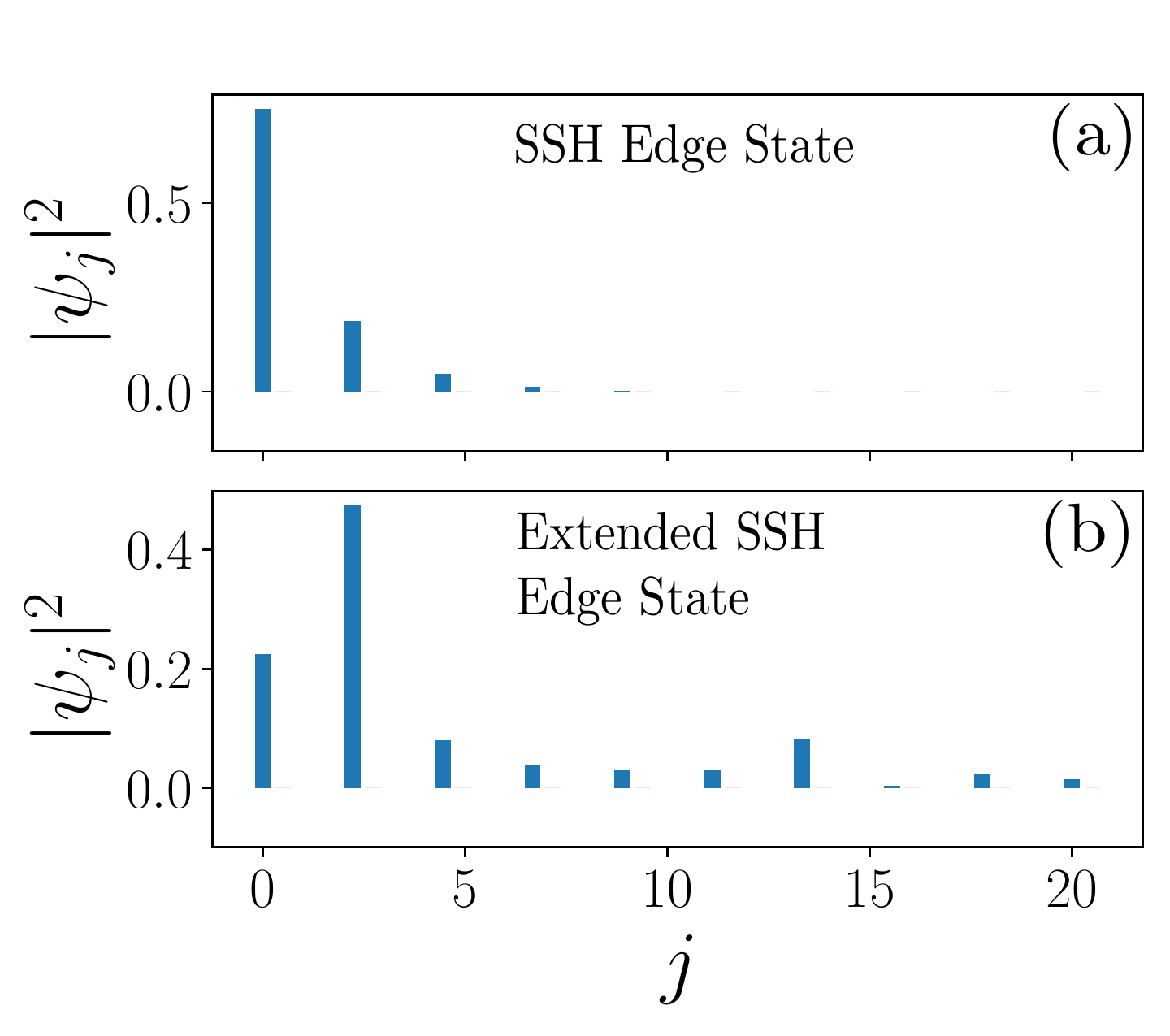}
\caption{Spatial structure of left edge-localized probability distribution of zero-energy modes in a) the SSH model with $J_1^\prime=2J_1$ and b) the extended SSH model with third-neighbour hoppings with parameters $(J_1^\prime,J_1, J_3^\prime,J_3)=(1,1,0.5,0.8)$. In both cases, the lattice size is of 20 sites (10 unit cells) labelled by $j$. Zero-energy eigenstates are hybridized combinations of states as the ones depicted in the figure. Blue bars depict support on chain sites belonging to sublattice A. No support in sublattice B for this edge state appears due to the chiral symmetry that both models display. }
\label{edge_states}
\end{figure}

The main interest of this manuscript is to understand the quantum optical consequences of coupling to topological baths with large winding numbers. For this reason, unless stated otherwise, we will consider the extended SSH model described above as the photonic bath. For concreteness, we will restrict the number of free parameters of the model by fixing $J_1=J_1^\prime=J$, and letting $J_3\neq J_3^\prime$. With this choice of parameters, the model displays the phase diagram shown in Fig.~\ref{phase_diagram}(b), with phases $\mathcal{W}=2,\pm 1,0$ depending on the relative value of $J_3/J,J^\prime_3/J$. As explained above, the energy spectrum displays two symmetric bands around the origin with a band-gap that depends on the tunneling difference $|J_3-J^\prime_3|$. For $J^{(\prime)}_3\ll J$, the energy spectrum features a very similar shape than the standard SSH model (see upper panel of Fig.~\ref{phase_diagram}(c)), with two middle band-edges around $k=\pm \pi$. Remarkably, when $J^{(\prime)}_3\gg J$ the bands acquire a qualitatively different shape (see lower panel of Fig.~\ref{phase_diagram}(c)). In particular,  bands are nearly periodic within the first Brillouin zone with period $2\pi/3$, i.e., the triple of its SSH-like counterpart, which leads to the appearance of new local maxima/minima within the bands. As we will see in the next sections, this new maxima/minima have important consequences when emitters couple to these type of baths.

The richer bulk topology that the extended SSH model displays in comparison with its nearest-neighbour counterpart also has implications on its boundary physics. As mentioned before, the bulk-edge correspondence~\cite{Ryu2002} links the absolute value of the topological invariant $ |\mathcal{W}|$ to the number of pairs of modes localized at the edges of the system, whose energy lies in the band-gap. Therefore, in a finite lattice with open boundary conditions, the spectrum includes $|\mathcal{W}|$ pairs of symmetric and antisymmetric combinations of these states localized at both ends of the lattice, with zero-energy in the thermodynamic limit. Apart from the larger number of edge modes in the $(\mathcal{W}=2)$-phase, another significant difference of the extended SSH model appears in their spatial shape. As shown in Fig.~\ref{edge_states}(a), the edge states of the standard SSH model both localize in a single sublattice of the model, and their amplitude decays exponentially as we look further away from the edge. The edge states of the extended SSH model~\cite{Li2014,Li2018,Maffei2018,Perez-Gonzalez2019}, on the contrary, display a non-monotonous decay of their wavefunction, although they still only have weight in one of the sublattices. In Fig.~\ref{edge_states}(b), we show an example of this for a particular edge state of the extended SSH model appearing in a phase with $\mathcal{W}=2$ (see parameters in the caption). The other edge states appearing in other phases show a qualitatively similar behaviour.

\subsection{Light-matter coupling~\label{sec:coupling_emitters}}

In this work, we are interested in analyzing what happens when emitters couple to a photonic environment described by the extended SSH model we have introduced in the previous section. Assuming the emitters have a single optical transition between an excited, $\ket{e}$ and a ground state, $\ket{g}$, the full light-matter Hamiltonian reads $\mathcal{H} = \mathcal{H}_\text{bath} + \mathcal{H}_\text{emitters}+\mathcal{H}_\text{int}$, where:
\begin{align}
\mathcal{H}_\text{emitters} =& \Delta \sum_\alpha \sigma^\alpha_{ee}\;,\\
\mathcal{H}_\text{int} =& g\sum_{\alpha} c_{j_\alpha}^\dagger \sigma^\alpha_{ge} +\text{H.c.}
\end{align}
where $\alpha$ is an index that runs over the set of emitters, $\Delta$ is the associated detuning of the emitter optical transition with respect to the reference energy $\omega_c$, $g$ is the emitter-bath coupling that we assume to the equal for all emitters, and $j_\alpha$ denotes the unit cell position the $\alpha$-emitter couples to. Here, $\sigma^\alpha_{ab} = |a\rangle_\alpha\langle b|_\alpha$, and $c_{j_\alpha}^\dagger$ can be either $a_{j_\alpha}^\dagger$ or $b_{j_\alpha}^\dagger$ depending on which sublattice the emitter couples to.

\begin{figure}[tb]
\includegraphics[width=\columnwidth]{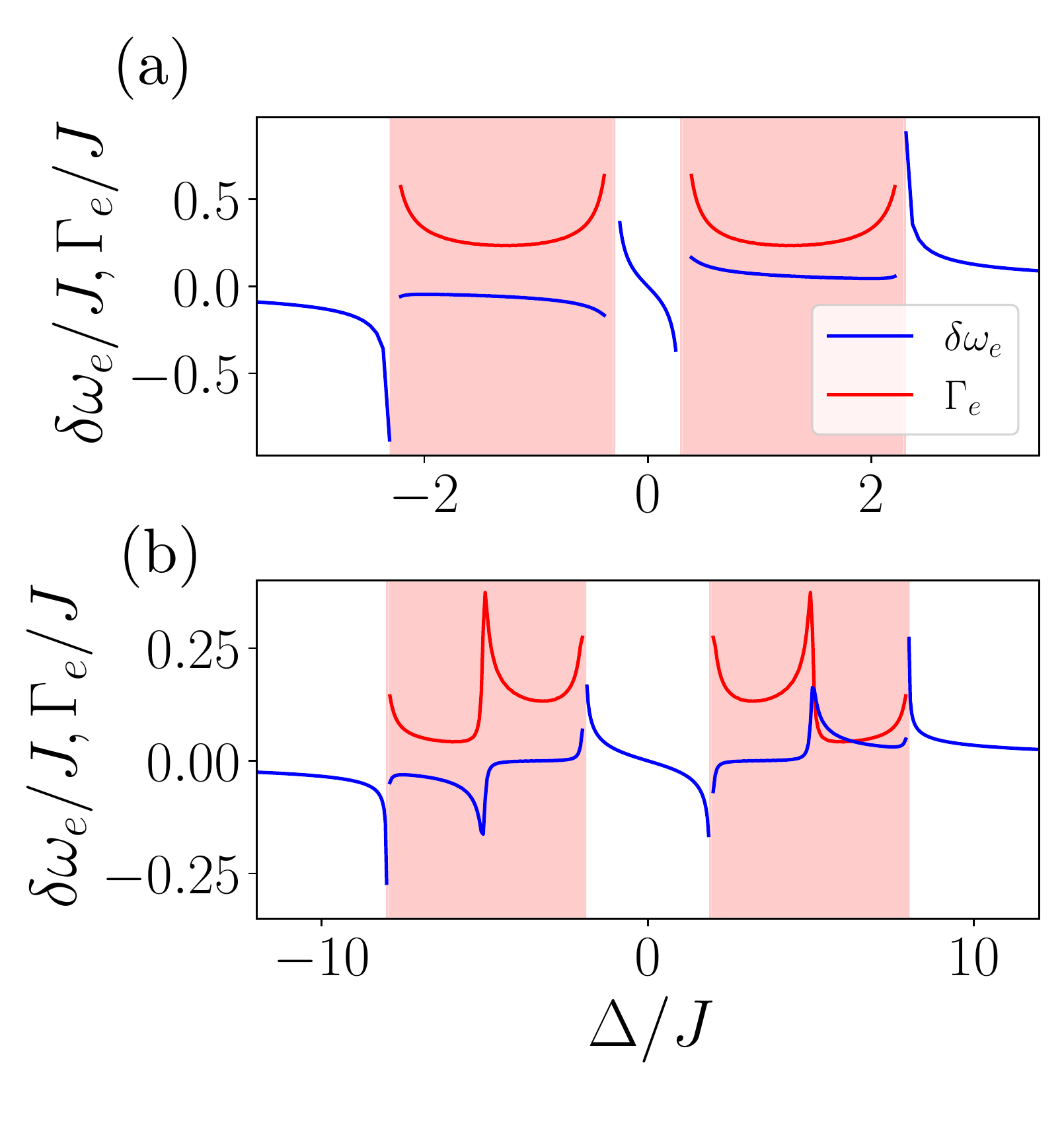}
\caption{Lamb shift $\delta\omega_e$ (blue lines) and Markovian decay rate $\Gamma_e$ (red lines) for a single emitter coupled with $g=0.5J$ to the $A$ sublattice of the extended SSH photonic bath with hoppings a) $(J_3^\prime, J_3)=(0,0.3)J$ and b) $(J_3^\prime, J_3)=(1,1,2,4)J$. Domains shaded red depict that $\Delta$ lies in the band regime.}
\label{self-energy}
\end{figure}

The coupling of emitters to an structured photonic environments, like the extended SSH one, strongly renormalizes their energies and lifetimes. A magnitude that captures these renormalizing effects is the single emitter self-energy, $\Sigma_e(z)$, a complex variable function which form, for a two-band model like the one we consider in this manuscript, reads~\cite{CohenTannoudji1998}:
\begin{equation}
\Sigma_e(z) = \sum_{k\in\text{BZ}} \frac{|\langle 0|u_k\mathcal{H}_\text{int}\sigma^\dagger|0\rangle|^2}{z-\omega_u(k)} + \frac{|\langle 0|l_k\mathcal{H}_\text{int}\sigma^\dagger|0\rangle|^2}{z-\omega_l(k)}\;.
\label{self_energy_eq1}
\end{equation}
where $u_k/l_k$ are the eigenoperators associated to the upper/lower band, $\omega_{u/l}(k)$, respectively, and $k$ runs over the first Brillouin zone (BZ) of the crystal, that is, $k\in[-\pi,\pi)$. In a one-dimensional chiral-symmetric bath, $|\langle 0|u_k\mathcal{H}_\text{int}\sigma^\dagger|0\rangle|^2 = |\langle 0|l_k\mathcal{H}_\text{int}\sigma^\dagger|0\rangle|^2=1/2$ and $\omega_{u,d}(k)=\pm\omega(k)$. Thus, the self-energy for a single emitter can be computed in the thermodynamic limit as
\begin{equation}
\Sigma_e(z)=\frac{g^2}{2\pi}\int_\text{BZ} dk\;\frac{z}{z^2-\omega^2(k)}\;.
\label{self_energy_eq2}
\end{equation}
The real and imaginary parts of $\Sigma_e(\Delta+i0^+)=\delta\omega_e(\Delta)-i\Gamma_e(\Delta)/2$ correspond to the Lamb shift $\delta\omega_e$, that renormalizes the emitter energy, and the Markovian decay rate $\Gamma_e$, that determines its lifetime in the perturbative regime. In Fig.~\ref{self-energy}, we plot both quantities, $\delta\omega_e(\Delta)$ and $\Gamma_e(\Delta)$, in blue and red, respectively, for an emitter coupled to the A sublattice and for an energy range $\Delta$ that spans beyond the two energy bands. In Fig.~\ref{self-energy}(a), we plot these magnitudes for the same scenario than the upper panel of Fig.~\ref{phase_diagram}(c), that is, when $J^{(\prime)}_3\ll J$. There, we observe how the Lamb shifts and lifetimes are very similar to the ones found in the standard SSH model~\cite{Bello2019}, with Van-Hove singularities at the band-edges for both the Lamb-shift and decay rates. On the contrary, for the regime when $J^{(\prime)}_3\gg J$, that we plot in Fig.~\ref{self-energy}(b), one observes an important qualitative difference, that is, the decay rate features an additional Van-Hove singularity in the middle of the upper/lower bands. The origin of these singularities is the presence of new local maxima of $\omega(k)$ that we show in Fig.~\ref{phase_diagram}(c). In those maxima, the group velocity, $v_g$, vanishes which leads to a divergence of the bath density of states since it scales inversely proportional to $v_g$. Such middle-band Van-Hove singularities are known to appear in two-dimensional structured reservoirs~\cite{Gonzalez-Tudela2017a,Gonzalez-Tudela2017b}, however, they are very unusual in one-dimensional baths. In Appendix~\ref{sec:emitter_band_dynamics}, we show how these Van-Hove singularities can have important dynamical consequences, such as the appearance of strongly non-Markovian dynamics of a single emitter with energies within the band, as it also occurs in higher dimensions~\cite{Gonzalez-Tudela2017a,Gonzalez-Tudela2017b}. In the main text, however, we will focus more on what occurs in the opposite regime, that is, when the energy of the emitter matches one of the three band-gaps appearing in Figs.~\ref{phase_diagram}(c),\ref{self-energy}.

\section{Emitters coupled to the photonic bulk: qubit-photon bound states~\label{sec:bulk}}

If an excited emitter is tuned off-resonance to the band structure of the photonic bath, it will not have available density of states to radiate its excitation. Thus, in this situation, the photon can not escape and it dresses the emitter, forming what is known as qubit-photon bound state~\cite{bykov75a,john90a,kurizki90a}. The energy and the wavefunction of this bound state can be calculated by solving the time-independent Schrödinger equation $\mathcal{H}|\psi_\text{BS}\rangle = E_\text{BS}|\psi_\text{BS}\rangle$ within the single-excitation subspace since $\mathcal{H}$ conserves the number of excitations. In this subspace, $|\psi_\text{BS}\rangle$ can be written as a superposition between an emitter and photonic excitation written as:
\begin{align}
|\psi_\text{BS}\rangle &=\ket{e}\ket{\mathrm{vac}}+\ket{g}\ket{\psi_\mathrm{ph}}=\nonumber\\
&= \left(C_e\sigma_{ge}+\sum_j C_{j,a}a_j^\dagger+C_{j,b}b_j^\dagger\right)|g\rangle|\text{vac}\rangle;.\label{eq:BSansatz}
\end{align} 

Solving formally this equation, one can find that the energy of the bound-state is given by the solution to the pole equation $E_\text{BS} - \Delta - \Sigma_e(E_{\text{BS}})=0$, with $E_{\text{BS}}$ outside of the band-gap. The wavefunction coefficients $C_{j,a}$ and $C_{j,b}$ can also be analytically obtained using this ansatz. For example, for an emitter coupled to the A sublattice at the $j=0$ unit cell, these coefficients read~\cite{Bello2019}:
\begin{align}
\frac{C^A_{j,a}}{C_e} =& \frac{gE_\text{BS}}{2\pi}\int_{-\pi}^{\pi}dk\;\frac{e^{ikj}}{E_\text{BS}^2-\omega^2(k)} \label{eq:bsa}\\
\frac{C^A_{j,b}}{C_e} =& -\frac{g}{2\pi}\int_{-\pi}^{\pi}dk\;\frac{\omega(k)e^{i(kj-\phi(k))}}{E_\text{BS}^2-\omega^2(k)}\;, \label{eq:bsb}
\end{align}
where $\phi(k)=\text{arg}\left(-J_1^\prime-J_1e^{-ik}-J_3e^{ik}-J_3^\prime e^{-2ik}\right)$. These coefficients are of particular interest since they
determine the shape of the 
emitter-emitter interactions when many emitters couple to the photonic bath~\cite{douglas15a,Gonzalez-Tudela2015b}. In particular, it has been shown that under the Born-Markov conditions, the effective emitter dynamics is governed by a purely Hamiltonian evolution given by~\cite{Bello2019}:
\begin{align}
H_{\mathrm{spin}}=\sum_{\alpha,\beta} J_{\alpha\beta} \left(\sigma_{eg}^\alpha\sigma_{ge}^\beta+\mathrm{H.c.}\right)\,,   \label{eq:heff}
\end{align}
where $J_{\alpha\beta}\propto C^A_{j,a}$ if both emitters couple to the same sublattice, and $J_{\alpha\beta}\propto C^A_{j,b}$ mode if they couple to different sublattices. Thus, the spatial shape of these single-qubit photon bound states is what ultimately determines how the emitters interact in the many-body regime.

For this reason, in this section we will focus on the study of the qubit-photon bound state properties of a single emitter coupled to the extended SSH bath. In particular, we investigate how the topological properties of the bath induce non-trivial features on these bound states, ranging from chiral spatial shapes~(\ref{sec:spatial_features}) to robustness against disorder~(\ref{sec:robustness}). We will probe the key role of topology in the bound states robustness by comparing these disorder effects with the ones in a topologically trivial model. Finally, in~\ref{sec:giant_atoms}, we explain how to further tune the spatial shape of such bound states by the use of non-local couplings, i.e. emitters coupled simultaneously to various lattice sites, that can be obtained using giant atoms~\cite{kockum18a,FriskKockum2021,Kannan2020,Wang2021a}.

\subsection{Spatial features~\label{sec:spatial_features}}

We will start the analysis of qubit-photon bound states describing their spatial shape when a single emitter is coupled to a bulk site of the photonic bath. We will assume that the emitter is coupled to the sublattice-A site at the $j=0$ unit cell. As we show in Fig.~\ref{phase_diagram}(c), since the structured bath is a two-band model, there are always three different band-gap regions in which bound-states can appear: the upper band-gap ($E_\mathrm{BS}>\text{max}\left(|\omega(k)|\right)$), the middle band-gap ($\text{min}\left(|\omega(k)|\right)>E_\mathrm{BS}>-\text{min}\left(|\omega(k)|\right)$) and the lower band-gap ($E_\mathrm{BS}<-\text{max}\left(|\omega(k)|\right)$). The upper and lower band-gap regions are also denominated outer band-gaps, to distinguish them from the topologically non-trivial band-gap that appears in the middle region. 

\begin{figure}[tb]
\includegraphics[width=\columnwidth]{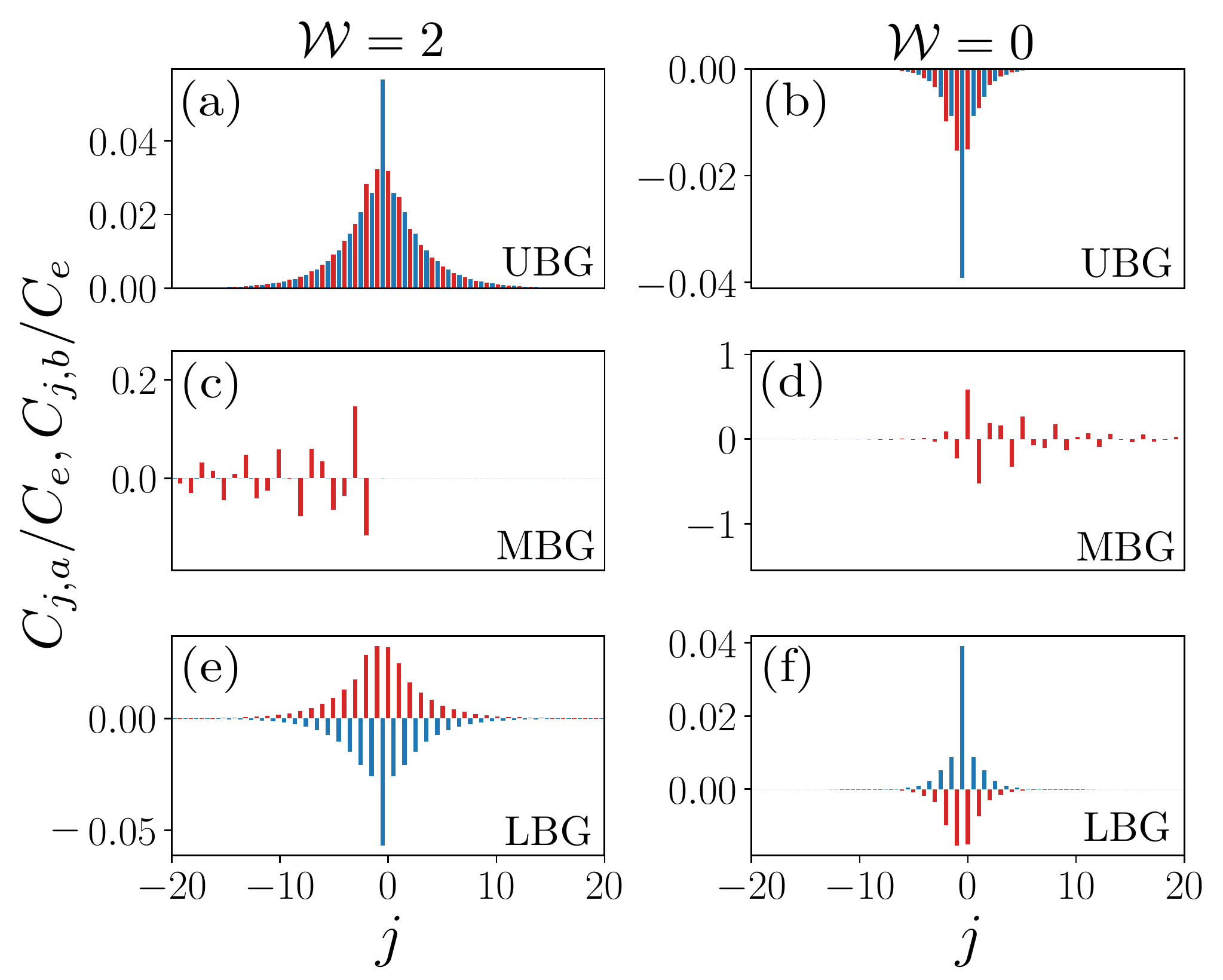}
\caption{Examples of spatial distribution of qubit-photon bound states around the emitter (coupled at position $j=0$) in a realization of the extended SSH model with $g/J=0.1$ for the upper band-gap (UBG) $\Delta/J = 3.5$ (a) and b)), middle band-gap (MBG) $\Delta/J=0$ (c) and d))) and lower band-gap (LBG) regimes $\Delta/J=-3.5$ (e) and f))) in a lattice of $N=600$ sites. Different topological phases are labeled by the winding number indicated at each sub-figure. Long-range hopping parameters have been chosen to have an equal photonic band-gap width $E_g=0.2J$ at each topological phase. Specifically, $(J_3^\prime,J_3)=(0.5J, 0.8J)$ in $\mathcal{W}=2$ and $(J_3^\prime,J_3)=(0.2661J, 0.5J)$ in $\mathcal{W}=0$. Bars coloured red (blue) represent support in sublattice B (A).}
\label{shape_bound_states_diff_detunings}
\end{figure}

In what follows, we analyze the bound-states of the model in the different phases, but noting first that changing $J_n\leftrightarrow J^\prime_n$ ($n=1,3$) only implies interchanging the role of the A/B sublattices. Thus, it suffices to study the topological phases with $\mathcal{W}=2,0$ (or $\mathcal{W}=-1,1$) to know the properties of bound states in every other phase. For this reason, in Fig.~\ref{shape_bound_states_diff_detunings} we only show the extended SSH bound state coefficients $C_{j,a}/C_e, C_{j,b}/C_e$ appearing for an emitter coupled to the A sublattice with $\Delta$ chosen in the different band-gaps (rows) for the $(\mathcal{W}=2,0)$-phases (columns). There, we observe how the bound states in the upper and lower band-gaps (panels (a-b) and (e-f), respectively) are exponentially localized to both sides of the emitter, and the only difference between them is a staggered sign between sublattices appearing in the lower band-gap case. This behaviour is very similar to the one found in standard SSH models, see Ref.~\cite{Bello2019}, and even in topologically trivial bandgaps. This is why we do not observe qualitative differences in the behaviour between the two distinct topological phases, $\mathcal{W}=2,0$, for these bound states. On the contrary, when the emitter matches exactly the central point of the middle band-gap, $\Delta=0$, the bound states are qualitatively different: they localize preferentially to one of the sides of the emitters for both phases, and have only weight in the opposite sublattice the emitter is coupled to. These two features also appear for the standard SSH bath~\cite{Bello2019}. However, in the extended SSH bath bound states also have their own qualitative differences: first, they feature an enlarged spatial periodicity as compared with the ones appearing in the standard SSH situation. 
Second, while in the $\mathcal{W}=2  \,(-1)$ phase the photon localizes perfectly at the left (right) of the emitter, i.e., is perfectly chiral, the one of the $\mathcal{W}=0\, (1)$ shows a small leakage to the other side. Third, they show a non-monotonous decay, like it occurred for the edge states of finite chains (see Fig.~\ref{edge_states}(b)). All these features can be understood by particularizing Eqs.~\eqref{eq:bsa} and~\eqref{eq:bsb} to the case $E_\mathrm{BS} = 0$, where we we have that
\begin{equation}
    C^A_{j,a} = 0 \,, \quad
    C^A_{j,b} \propto \frac{1}{2\pi} \int dk\, \frac{e^{ikj}}{f(e^{ik})} \,,
\end{equation}
with $f(e^{ik}) = J^\prime_3 e^{ik} + J^\prime_1 + J_1 e^{-ik} + J_3 e^{-2ik}$. Doing a change of variable $y = \exp[\text{sign}(j)ik]$, we can transform the integral into a contour one along the unit circle that can be solved using Residue Theorem finding the poles of $1/f(y)$ (see Appendix~\ref{sec:BS_shape} for details). In fact, one can show that the number of poles within the unit circle is given by $2-\mathcal{W}$, which explains why in the $\mathcal{W}=2,-1$ phases the bound state shows a perfect multi-exponential localization just one side of the emitter, whereas in the $\mathcal{W}=0,1$ phases it is localized on both sides, with a non-monotonous decay only at one side.

As aforementioned this non-monotonous decay is reminiscent of the edge-state shape appearing in finite chains (see Fig.~\ref{edge_states}(b)). This points to the topological origin of the qubit-photon bound states in these baths. In the SSH case, this connection was formally proved by noticing that the photonic component of the bound-states is precisely the topological edge state appearing in one of uncoupled semi-infinite chains that appears when considering a vacancy defect in the position of the bath the emitter couples to~\cite{Leonforte2020b}. In the extended SSH model, this connection is not so obvious since the long-range hoppings connect the two semi-infinite chains and their modes become hybridized. However, it can also be shown that in the extended SSH case, the photonic component of the bound states can be written as linear superpositions of the edge states ($\ket{\mathrm{ES}_j}$) of the two semi-infinite chains that are obtained when breaking the lattice at the position of the emitter~\cite{Leonforte2020b}, i.e., $\ket{\psi_\text{ph}}= \sum_j\braket{\text{ES}_j|\psi_\text{ph}}\ket{\text{ES}_j}$ (see Appendix~\ref{sec:BS_shape} for explicit formulas). Thus, this observation strongly suggests that qubit-photon bound states will inherit topological protection even when long-range hoppings are included in the photonic environment. This is what we study in more detail in the next section.

\subsection{Robustness to disorder~\label{sec:robustness}}

One of the most remarkable properties of topology in both condensed-matter and photonic systems is the protection against disorder of gapless boundary modes that emerge when the bulk topology is non-trivial. In this section, we explore whether this protection is inherited by the energy and shape of the qubit-photon bound states when an emitter couples to the extended SSH model. 

To analyze this robustness, we will consider two different types of disorder in the bath Hamiltonian, depending on whether they preserve the chiral symmetry of the model or not. For chiral-preserving disorder, we consider the addition of random perturbations to the hopping amplitudes that the model already incorporates (first and third-neighbour hoppings), which do not change the BDI topological class of the hamiltonian. Mathematically, chiral-preserving disorder is implemented as:
\begin{equation}
\mathcal{H}\rightarrow\mathcal{H}+\sum_{|i-j|=1} \varepsilon^{i,j}_1 c_i^\dagger c_j + \sum_{|i-j|=3} \varepsilon^{i,j}_3 c_i^\dagger c_j\;,
\end{equation}
where the $|i-j|=N$ subscript depicts a pairs of neighbouring sites at distance $N$, and $c_i$ is either $a_i$ or $b_i$ depending on the sublattice that the site belongs to. The coefficients $\varepsilon^{ij}_1$ and $\varepsilon^{ij}_3$ are random variables described by a certain probability distribution that we will take to be gaussian with zero mean $\mathcal{N}(0,\sigma)$. The standard deviation $\sigma$ of the distribution acts as the strength of the induced disorder. On the other hand, chiral-breaking disorder will include random diagonal terms and second neighbour hoppings, which turns the topological class of the model from BDI to AI. These disorder terms read:
\begin{equation}
\mathcal{H}\rightarrow\mathcal{H}+\sum_{i} \varepsilon^{ii}_0 c_i^\dagger c_i + \sum_{|i-j|=2} \varepsilon^{i,j}_2 c_i^\dagger c_j\;.
\end{equation}

\begin{figure}[tb]
\includegraphics[width=\columnwidth]{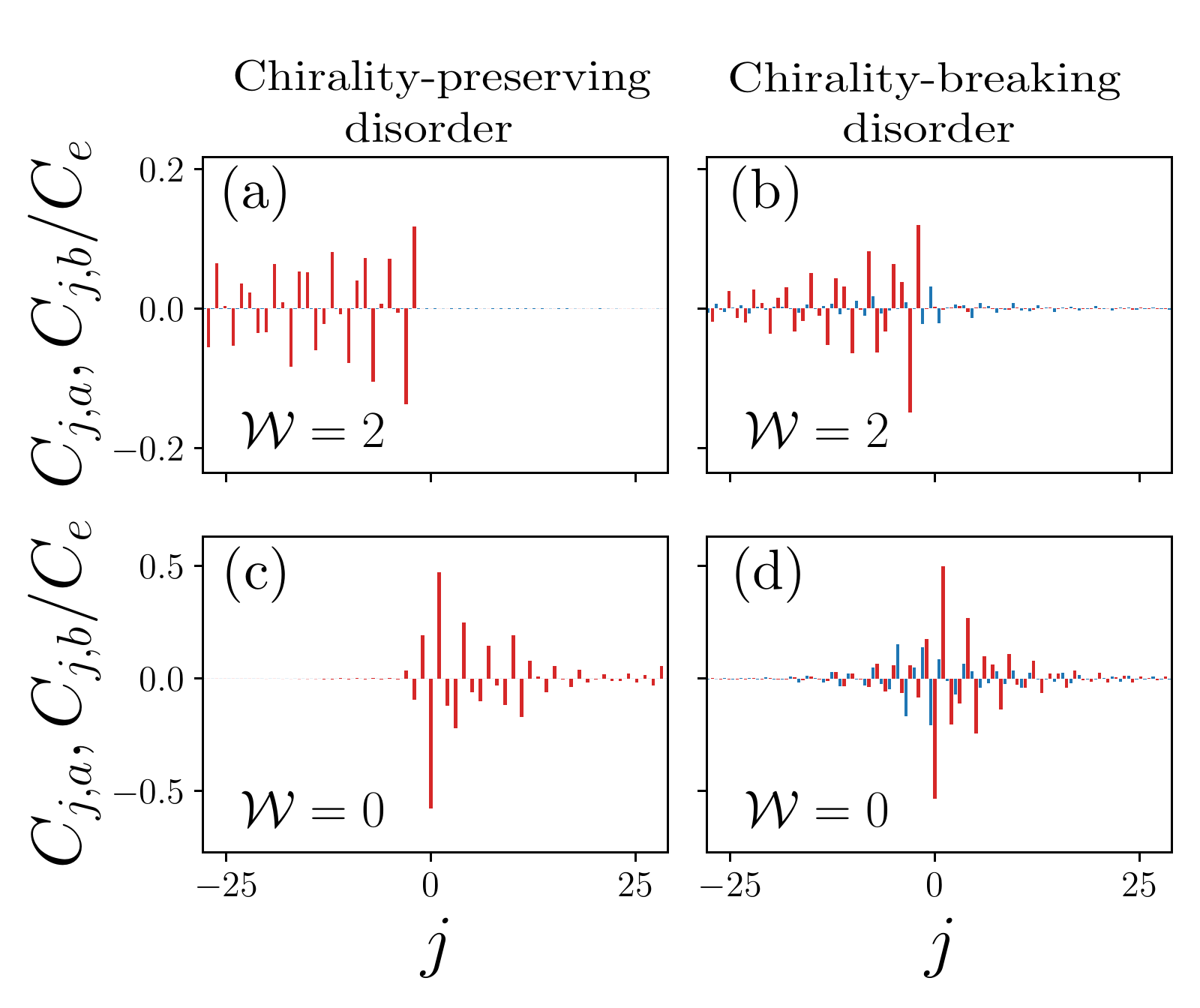}
\caption{Effects of different chirality-preserving (a) and c)) and chirality-breaking (b) and d)) types of disorder on the bound states wavefunctions with a emitter detuning of $\Delta/J=0$ in a lattice of $N=600$ sites. In all cases, the disorder strength is $\sigma/J=0.1$. Chirality-breaking disorder manifestly induces support of the bound state in both sublattices. Hopping amplitudes are set to $(J_3^\prime,J_3)=(0.5J, 0.8J)$ in $\mathcal{W}=2$ and $(J_3^\prime,J_3)=(0.2661J, 0.5J)$ in $\mathcal{W}=0$.}
\label{disorder}
\end{figure}

Robustness against disorder manifests as the protection of zero-energy modes localized around generic defects or boundaries (in our case, a emitter) of a system with non-trivial bulk topology. Provided that disorder terms preserve the symmetries that characterize the topological class of the model, and that its strength is small enough (less than the energy difference between the topological state and the closest band energy), these topological modes should persist if such random perturbations are added to the bath Hamiltonian. In our case, it is then expected that only under chiral-preserving disorder, qubit-photon bound states will remain protected. In Fig.~\ref{disorder} we show several examples of the effects of disorder that preserves and breaks chiral symmetry in the different columns, and for two distinct phases ($\mathcal{W}=0,2$) in the different rows. For chiral-preserving disorder, the bound-states maintains qualitatively their shape: still have support in one of the sublattices, and are localized preferentially to either the left/right depending on the phase considered. As expected, the effect of chirally-breaking disorder is more dramatic: the bound-states acquire weight in the other sublattice, and in the $(\mathcal{W}=0)$-phase loses significantly its chiral character.

Beyond the spatial features of the bound states, disorder is also expected to affect their energies. When randomly sampling disordered configurations of the bath hamiltonian, the bound state energy $E_\text{BS}$ is calculated by diagonalizing the complete light-matter Hamiltonian, $\mathcal{H}$, and searching for the localized eigenstate. When doing that for a fixed disorder strength and several disorder realizations, the energy of the bound-state is a random variable with some probability distribution with mean $\langle E_\text{BS}\rangle$ and standard deviation Std($E_\text{BS}$). In what follows, we investigate the evolution of $\langle E_\text{BS}\rangle$ and Std($E_\text{BS}$) as the disorder strength $\sigma$ increases in several scenarios. First, we will compare the protection of the middle band-gap bound states of the extended SSH bath model to the ones of the model with a topological trivial middle band-gap obtained by introducing the staggered cavity energy shifts $\omega_\delta$ (see Section~\ref{sec:system}). The motivation for the comparison is to discern whether the bipartite nature of the bath, which is what ultimately leads to the middle band-gap opening, plays a role in the "protection" of the bound-states or it is really the topological nature of the bath what makes a difference. Then, we will also compare the resilience to disorder of the bound-states appearing in the outer band-gaps, to see if this resilience extends also to those situations.

\begin{figure}[tb]
\includegraphics[width=\columnwidth]{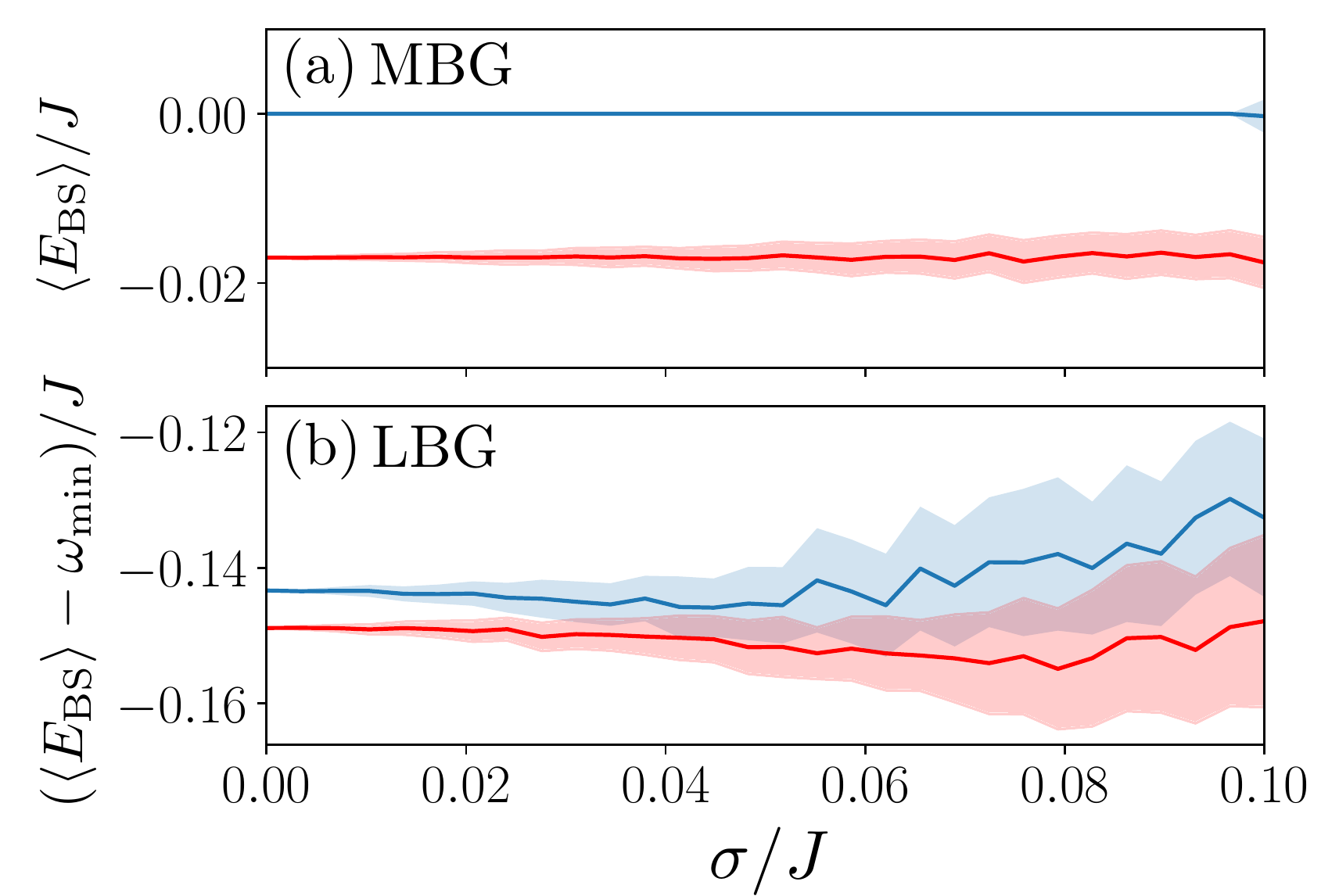}
\caption{Evolution of mean bound state energies $\langle E_\text{BS}\rangle$ for an extended SSH (blue lines) lattice a and staggered standard waveguide (red line). a) The means of bound states are computed in the a) middle band-gap (MBG) ($\Delta/J=0$) and b) lower band-gap (LBG) regime, with a detuning given by $\Delta=\omega_\text{min}-E_g/2$ (set to $\Delta/J=-3.3$ for the extended SSH and $\Delta/J=-2.1$ for staggered standard waveguide). In the LBG case, $\langle E_\text{BS}\rangle$ is expressed for both models referred to the minimum band energy $\omega_\text{min}$. The mean energies are computed among samples of $200$ realizations of chirality-preserving disorder with fixed $\sigma$ in lattices of $600$ sites, for a coupling constant $g=0.2J$. The extended SSH model hoppings are $(J_3^\prime, J_3)=(0.5J,0.8J)$, with band-width $E_g=0.21J$, while the staggered mass in the standard waveguide model is set to yield equal band-gap width $E_g$. Shaded regions surrounding solid lines depict the standard deviation around the mean value.}
\label{disordered_BS_energy}
\end{figure}

To begin this analysis, in Fig.~\ref{disordered_BS_energy} we plot $\langle E_\text{BS}\rangle$ (in solid lines) and their Std($E_\text{BS}$) (in shadow) for the extended SSH (blue) and staggered energy models (red), and for bound-states appearing in the middle/lower bound-states for the (a)/(b) panels, respectively. To make a fair comparison, we fix the parameters of the two models so that they feature the same middle band-gap width $E_g$ (see the caption for parameters). From this figure, we can extract several conclusions: first, we observe how the middle band-gap bound states of the extended SSH model (blue lines) are pinned at zero for $\Delta=0$ for a wider range of disorder strengths. Only when $\sigma\sim E_g$, the Std($E_\text{BS}$) of such middle bound-states acquire a significant value. This is in stark contrast to the bound-states of the topologically-trivial model (in red) where we observe that: the mean value $\langle E_\text{BS}\rangle$ oscillates around a value different from zero, due to the chiral symmetry-breaking of the model, which makes the bound-state energies different from zero even if $\Delta=0$. Besides, the standard deviation Std($E_\text{BS}$) grows continuously for increasing values of disorder strength, unlike the topologically trivial case. From this, we can conclude that the bound-states energies of the extended SSH model are definitely more protected than for the other model. For completeness, in panel (b), we plot the behaviour of $\langle E_\text{BS}\rangle$ and their Std($E_\text{BS}$) for the case of bound-states in the lower band-gap, fixing $\Delta$ so that the detuning with the respect to the lower-band-edge is the same than in panel (b) to make a fair comparison. There, we observe that both models feature a similar qualitative behaviour: the mean value gets displaced as $\sigma$ increases and Std($E_\text{BS}$) grows continuously. This hints to a lack of topological protection of these bound-states, irrespective of the topological nature of the bath. We characterize this in more detail in Appendix~\ref{sec:disorder_driven}. 

\begin{figure}[tb]
\includegraphics[width=\columnwidth]{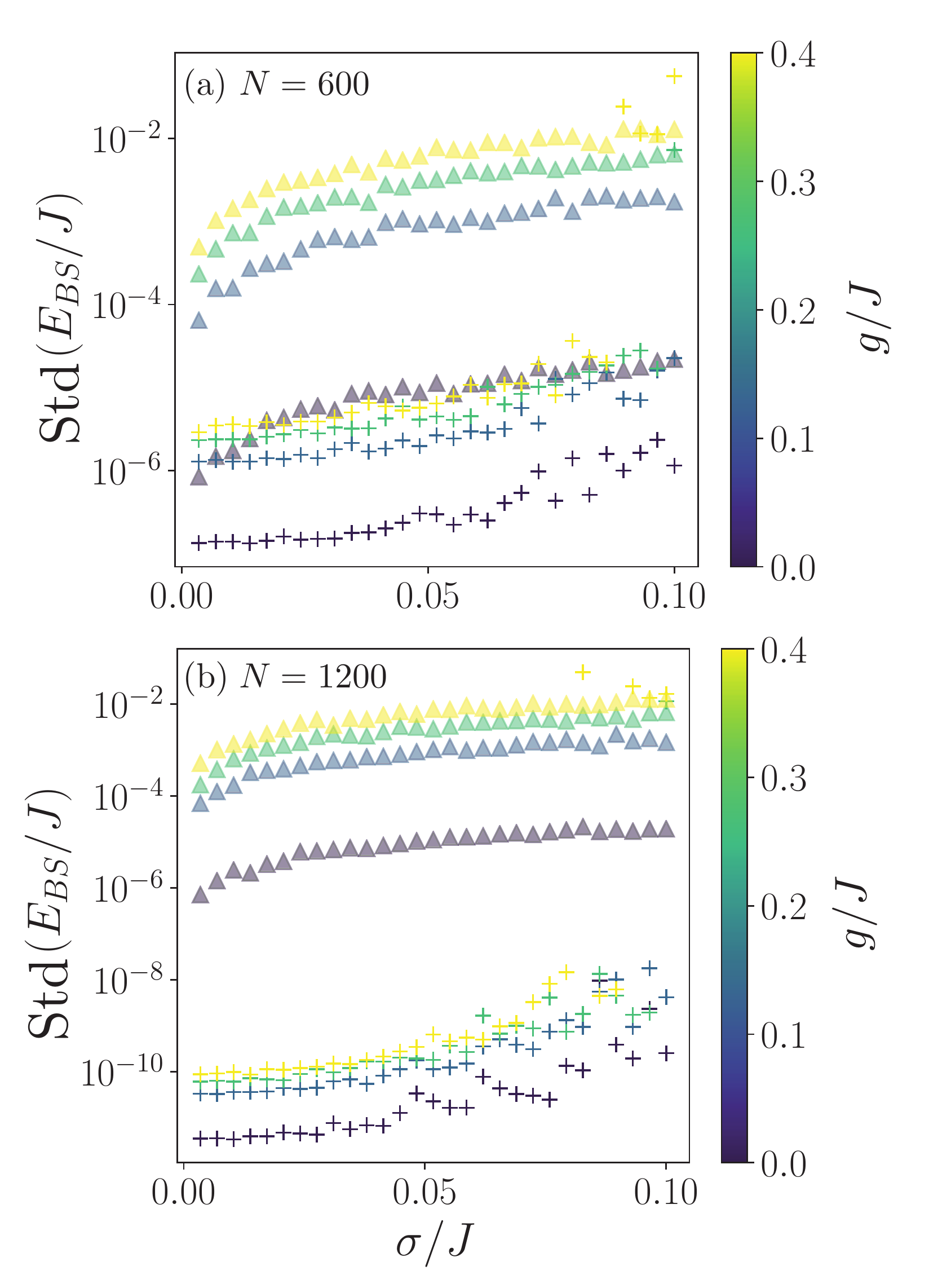}
\caption{Standard deviation of qubit-photon bound state energies $E_\text{BS}$ as gaussian disorder strength $\sigma$ is increased. Different values of the coupling constant $g$ are considered for a emitter detuning in the centre of the topological (crosses) and trivial (triangles) band-gaps, for lattices of a) $N=600$ sites and b) $N=1200$ sites. The parameters of both models have been set as in Fig.~\ref{bandgap_protection}. In both figures, each point corresponds to the standard deviation in a sample of $50$ realizations of chirality-preserving disorder with fixed $\sigma$.}
\label{bandgap_protection}
\end{figure}

The aforementioned differences in Std($E_\text{BS}$) suggest that middle band-gap bound states are indeed more robust to disorder in the topological model due to their topological origin. To further assess this statement, we now make a more detailed study on how Std($E_\text{BS}$) scales with two relevant parameters, namely, the coupling strength $g$ and the system size $N$. The coupling strength $g$ governs the amount of photonic component of the qubit-photon bound-states ($\sum_{j,\alpha=a,b}|C_{j,\alpha}|^2$ in Eq.~\ref{eq:BSansatz}), which can be shown to be proportional to $g^2/\Delta^2$ for this situation. Thus, since the photonic compoment of the topological qubit-photon bound states with $\Delta=0$ is built out of superpositions of the topological edge states, which are robust to disorder, we expect that Std($E_\text{BS}$) is much less dependent on the coupling strength $g$ than in the topologically trivial case. Besides, for the same reason, we expect that this topological protection increases with system size for the topological case, since it is known that topological edge states are perfectly insensitive to disorder at the thermodynamic limit~\cite{asboth15}.

In Figs.~\ref{bandgap_protection}(a-b) we numerically confirm these hypotheses by plotting the evolution of the Std($E_\text{BS}$) for increasing disorder strength $\sigma/J$, for both the topologically trivial (filled markers) and non-trivial models (crosses) of Fig.~\ref{disordered_BS_energy}. The different colors represent different values of the coupling strengths, $g$, as indicated in the right color bar, and the two panels correspond to different system sizes, namely, $N=600$ and $N=1200$ for panels (a) and (b), respectively. There, we can already see that for fixed system size and coupling strengths, the variances Std($E_\text{BS}$) of the topological bound states are always smaller that the topologically trivial ones. Besides, the dependence of Std($E_\text{BS}$) on the coupling strength is much weaker than in the other situations. Finally, comparing the two panels (a) and (b), we see how doubling the size of the system decreases dramatically the Std($E_\text{BS}$) in the topological case, while leaves almost unaltered the variances in the trivial situation. This suggest that the observed variances for the topological model are indeed a finite size effect, that vanishes in an infinite lattice regardless of the value of the coupling constant $g$. Finally, we wanted to note that although Fig.~\ref{bandgap_protection} depicts energy variances only for the extended SSH $(\mathcal{W}=2)$-phase, repeating the same numerical analysis in the other topological phases of the model we found the same qualitative behaviour. Thus, we conclude that middle band-gap qubit-photon bound states indeed inherit the topological protection from the bath.

\subsection{Tunability through non-local couplings~\label{sec:giant_atoms}}

\begin{figure}[tb]
   \includegraphics[width=0.99\columnwidth]{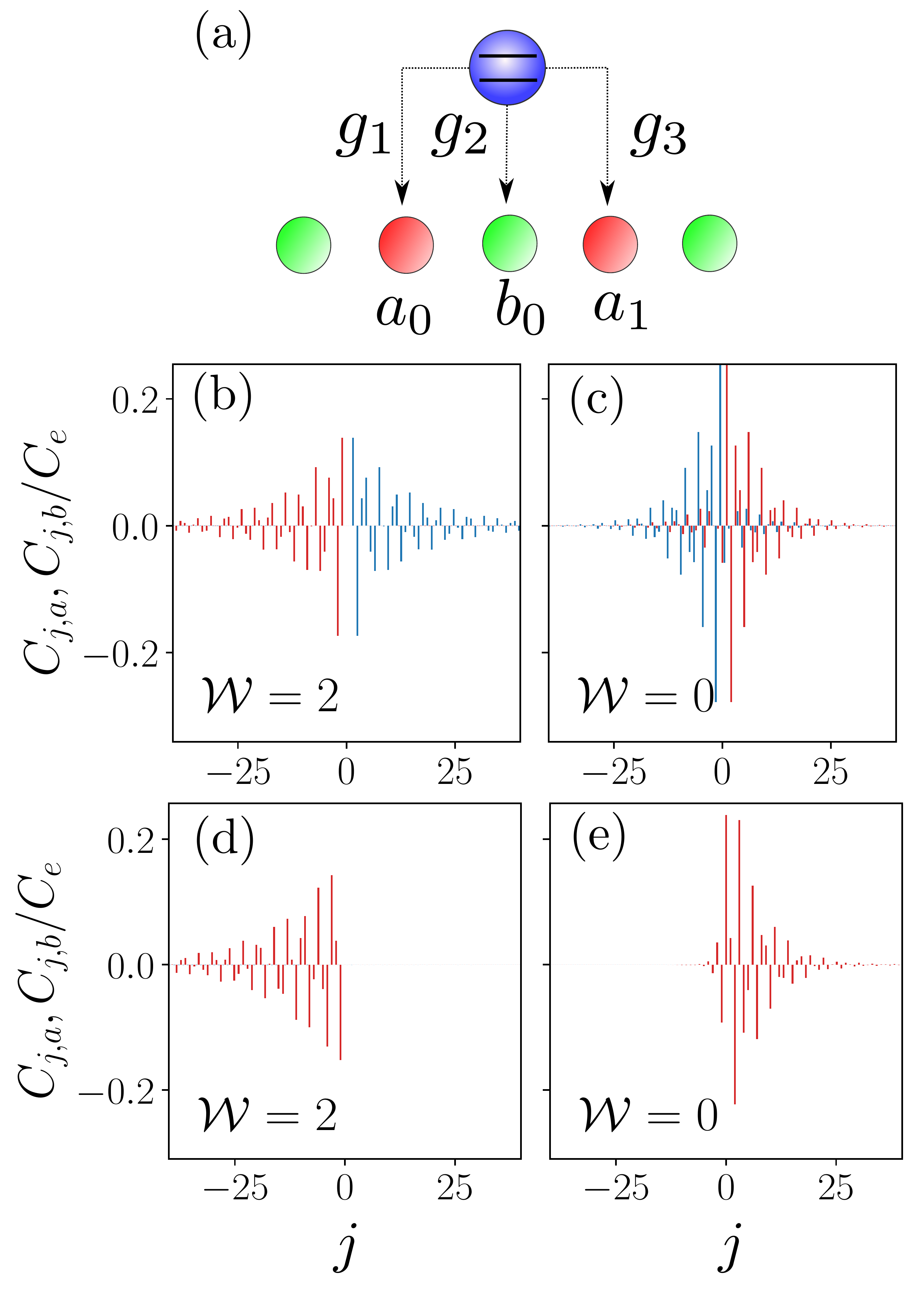}
  \caption{Qubit-photon bound states distribution when a giant atom with $\Delta=0$ is coupled to the lattice as depicted in the scheme a). The considered values for the coupling constants are $(g_1,g_2,g_3)=(g,g,0)$ (b) and c)), and $(g_1,g_2,g_3)=(g,0,g)$ (d) and e)), where $g=0.2J$. Each bound state spatial distribution corresponds to the topological phases characterized by $\mathcal{W}=2 $ and $\mathcal{W}=0$ indicated in each subfigure with hopping amplitudes given by $(J_3^\prime,J_3)=(0.5J, 0.8J)$ in $\mathcal{W}=2$ and $(J_3^\prime,J_3)=(0.2661J, 0.5J)$ in $\mathcal{W}=0$.} 
  \label{giant_atoms}
\end{figure}

After having discussed the spatial and energetic features of the qubit-photon bound states appearing in the extended SSH, here we aim to show how to tune their spatial shape through the use of non-local light-matter couplings. These type of couplings are motivated by recent experimental advances in circuit QED setups, where qubits can be made to interact with waveguides simultaneously at multiple points several wavelengths apart, forming what has been labelled as \emph{giant atom}~\cite{FriskKockum2021}. These multi-point couplings lead to strong interference effects which can be harnessed, for example, to engineer decoherence-free quantum gates~\cite{kockum18a}, directional emission~\cite{ramos16a,Gonzalez-Tudela2019b}, to probe topology through photon scattering~\cite{Bello2019}, or to induce tunable inter-emitter couplings~\cite{Wang2021a}, as we will do here.

For concreteness, we will restrict to the situation where the emitter couples to two different lattice sites. Since our bath is bipartite, there are two nonequivalent physical situations depending on whether it couples to the same or a different sublattice. For illustration, we consider each of these situations for two non-local emitter-bath coupling scenarios schematically depicted in Fig.~\ref{giant_atoms}(a): (i) when the emitter couples to both sites A and B of the central unit cell $j=0$, or (ii) when it couples to the A sites of two consecutive unit cells. Thus, the interaction Hamiltonian will be of the form 
\begin{equation}
    \mathcal{H}_\text{int} = \left(g_1 a_0^\dagger+g_2 b_0^\dagger+g_3a_1^\dagger\right)\sigma_{eg} + \text{H.c.}\;,
\end{equation}
with $(g_1,g_2,g_3)=(g,g,0)$ and $(g_1,g_2,g_3)=(g,0,g)$ for the (i) and (ii) situations. In Figs.~\ref{giant_atoms}(b-c) we plot the (middle band-gap) bound-state shape for the (i)-situation for the $(\mathcal{W}=2)$ and $(\mathcal{W}=0)$-phases, respectively. There, we observe a curious feature: the bound-state loses its chiral character, since the photonic component localizes at the both sides of the emitter. However, at each side it localizes preferentially in one of the sublattices (note the different color). For the (ii)-situation depicted in Figs.~\ref{giant_atoms}(d-e), respectively, the bound-state recovers its chiral shape for both topological phases. As it occurs in the local coupling case, the bound states in the ($\mathcal{W}=2$)-phase appear to exhibit a stronger chiral character than in the other situation.

\section{Emitters interacting with the edge modes of finite systems~\label{sec:edge}}

All the previous results considered situations where the emitter couples to the \emph{bulk} modes of the chain, such that the physics is not affected by the edge states appearing in finite systems whenever $\mathcal{W}\neq0$. In this section, however, we study precisely the opposite situation, that is, when the emitter couples precisely to the lattice sites at the end of the chain. One of the motivations for considering this configuration is to find some quantum optical observable sensitive to the number of edge states of the system, and thus, can be used as a probe of the topological phase of the bath. 

With this in mind, we consider the time evolution of a quantum emitter coupled to one of the edges of a finite extended SSH chain. If the emitter frequency is tuned to the middle bandgap, we expect the coupling to the topological edge states to be dominant, leading to qualitative different dynamics for each topological phase. Considering local couplings, however, we find that the emitter dynamics is not able to distinguish between all the different phases (see Fig.~\ref{dynamics_local_coupling} of Appendix~(\ref{sec:effective_model})). The reason is that, due to chiral symmetry, the edge states in some of the topological phases do not have support in the sublattice the emitter couples to, and thus, the emitter does not couple to them even if they are resonant. Therefore, the dynamics of a single emitter can not be sensitive to all the different phases regardless of the configuration. Interestingly, as we will show below there is a way of recovering the sensitivity to all phases by using the non-local couplings introduced in section~\ref{sec:giant_atoms}. 

\begin{figure}[tb]
\includegraphics[width=\columnwidth]{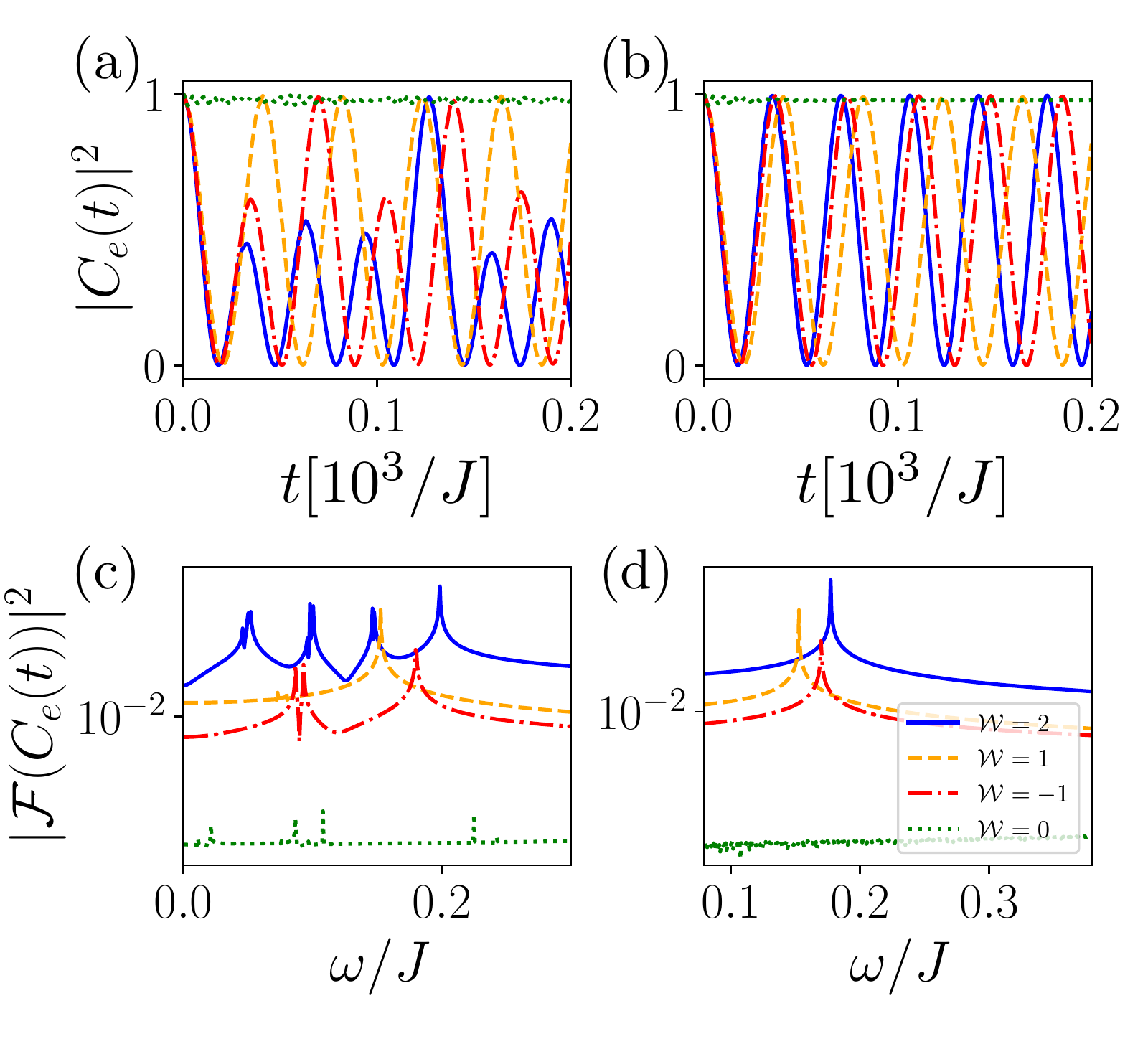}
\caption{Dynamics in time and frequency domains of a giant atom with $\Delta/J=0$ coupled to both sites of the leftmost unit cell of an extended SSH photonic crystal of $20$ sites (a) and c)) and of $120$ sites (b) and d)), with a coupling constant $g/J=0.1$. Hopping amplitudes are set to $(J_3^\prime, J_3)=(2J,0.5J)$ in $\mathcal{W}=-1$, $(J_3^\prime, J_3)=(0.5J,-0.76J)$ in $\mathcal{W}=1$, $(J_3^\prime, J_3)=(0.5J, 2J)$ in $\mathcal{W}=2$ and $(J_3^\prime, J_3)=(-0.76J,0.5J)$ in $\mathcal{W}=2$. All these parameter configurations lead to a band-gap width of $E_g = 2.21J$.}
\label{edge_dynamics_1}
\end{figure}

Let us illustrate this sensitivity considering a situation where the emitter couples simultaneously to both sublattices A and B of the leftmost unit cell of the lattice. In Figs.~(\ref{edge_dynamics_1}(a-b), we plot the emitter population dynamics, $|C_e(t)|^2$, for two different system sizes $N=20$ and $N=120$, respectively, and considering the different topological phases in different colors (see legend), whereas in Figs.~(\ref{edge_dynamics_1}(c-d) we plot their corresponding spectral components by making a Fourier analysis. For the smaller lattice, we clearly see very different dynamical features for all the different topological phases: the $(\mathcal{W}\neq 0)$-phases feature (multi)-frequency oscillations due to the coupling to the different edge-states, whereas the $(\mathcal{W}=0)$ one shows predominantly a no-decay dynamics since there are no modes energetically available to exchange interactions with. These multi-frequency coherent exchanges are more evident in the analysis of the spectral components of  Fig.~\ref{edge_dynamics_1}(c), where we observe different number of peaks for each topological phase, which can be understood from the hybridization of the emitter with the different edge states of the chain. Since the overlap between the different topological edge-states decreases exponentially with system size, we expect that these multi-frequency exchanges disappear for larger systems, as we show in Figs.~\ref{edge_dynamics_1}(b),(d). There, we observe how the only difference between the $\mathcal{W}\neq 0$ appears in the quantitative value of the single-frequency Rabi oscillation that appear in the emitter dynamics.  All these dynamical features can be reproduced within an effective model obtained by projecting into the edge state subspace, as we show in Appendix~(\ref{sec:effective_model}).

\section{Experimental implementation~\label{sec:exp}}

In this section, we propose a setup to implement our ideas with superconducting circuits, an experimental platform where all the elements needed for our model are already available. 
In fact, the coupling of qubits to coupled cavity arrays~\cite{liu17a,Sundaresan2019,Mirhosseini2019}, 
including one mimicking the standard SSH model~\cite{Kim2020b}, has been recently implemented. 
In the following, we will specifically focus on how to implement longer-range hoppings, which is the additional ingredient for the extended SSH Hamiltonian. 
Controlling long-range hopping terms is an issue that has been considered in the literature, see ~\cite{Onodera2020} and~\cite{Puri2017}, since this is essential, for example, for quantum annealing applications. 
Previous works have been mostly interested in achieving full connectivity between cavities or qubits by means of Floquet engineering, and those schemes could be used for implementing our model as well. Below we propose a more economic alternative that achieves just the connectivity that is necessary for our extended SSH model.

\begin{figure}[tb]
\includegraphics[width=\columnwidth]{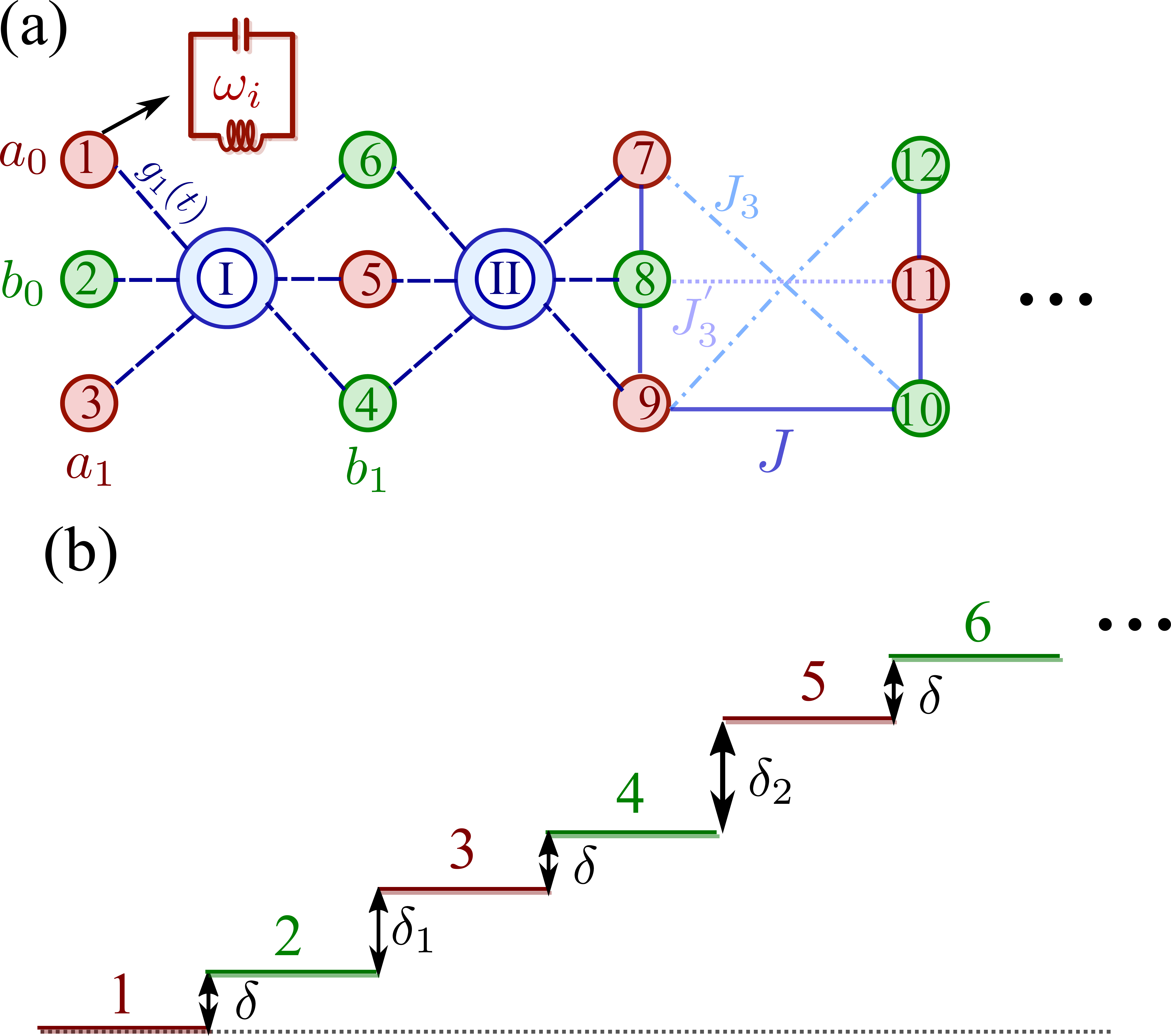}
\caption{(a) Scheme of a possible circuit QED arquitecture to implement the bath Hamiltonian: a set of coupled LC-resonators, with frequency, $\omega_i$, is coupled through fixed capacitive couplings, $J$, to their nearest neighbours. Red (green) refers to sublattice A (B). An index labelling each site is depicted along each cavity, which is within each cavity, Apart from that, the resonators are coupled in groups of six resonators to auxiliary resonators (in blue) with frequency $\omega_\mu$ and time-dependent couplings $g_\alpha(t)$. Using an appropriate choice of $\omega_i$ and $g_\alpha(t)$ tunable third-neighbour hoppings, $J_3^{(\prime)}$, can be engineered (as schematically depicted in the right of the figure. (b) Energy distribution of resonators, $\omega_i$, to implement the desired bath Hamiltonian. One must use an intra-cell energy difference $\delta$ and inter-cell, $\delta_i$.}
\label{circuit}
\end{figure}

The main challenge when implementing our extended SSH Hamiltonian is to have a coupled cavity array as the one depicted in Fig.~1(a), such that even hoppings vanish, while odd-ones do not. In order to do it, we propose a combination of fixed capacitive couplings for the nearest neighbour couplings $J$, that are fixed parameters of our model, plus the use of Floquet engineering with time modulated couplings, as already used experimentally for other purposes~\cite{Chen2014,Peropadre2013a,Roushan2017,Baust2015}, to implement the tunable third-neighbour hoppings, $J_3^{(\prime)}$. For that, we couple the six cavities extending along three unit cells to auxiliary cavities with frequency $\omega_\mathrm{\mu}$ by means of adjustable couplers $g_\alpha(t)$~\cite{Chen2014,Peropadre2013a,Roushan2017,Baust2015} (denoted by Roman numbers I, II, III...), 
as shown in Fig.~\ref{circuit}(a). 
These  auxiliary cavities will mediate long-range hopping terms, whose activation will be determined by resonances induced by periodic modulation of the adjustable couplers.

Let us show quantitatively how the scheme works.  
We assume that the main resonators have all different frequencies 
$\omega_{\alpha}$, $\alpha = 1, 2, \dots$,
something that will allow us to control the couplings.
For simplicity we consider that all the auxiliary cavities 
have the same frequency $\omega_\mu\equiv \omega_{\rm aux} $, for all $\mu$.
The interaction Hamiltonian that couples the main and auxiliary cavities is, in the interaction picture, 
\begin{equation}
\mathcal{H}_\text{aux} = 
\sum_{\mu = \mathrm{I}, \mathrm{II}, \dots } 
\sum_{\langle \alpha \rangle_\mu} g^{\mu}_\alpha (t) 
c_\alpha^\dagger b_\mu 
e^{i (\omega_\alpha - \omega_{\rm aux}) t}+\text{H.c.}\,,\label{eq:Hinttime}
\end{equation}

We need to work in a regime where the main cavities are far detuned from the auxiliary cavities, such that we can eliminate the coupling adiabatically. On the other hand, we assume that differences in frequencies between the main cavities are smaller than that detuning, but still larger than the couplings $g_{\alpha}^\mu(t)$, so that photon hopping is forbidden unless activated by periodic driving. 
This leads to the following hierarchy of frequencies, 
\begin{equation}
    \omega_\alpha - \omega_{\rm aux} \gg |\omega_\alpha - \omega_\beta| \gg g^\mu_{\alpha}(t).
    \label{eq:limits}
    \end{equation}

Within this regime, we can adiabatically eliminate the coupling between the auxiliary and main cavities, assuming that $g^\mu_\alpha(t)$ is real and varies slowly on the time scale of the inverse of $\omega_\alpha - \omega_{\rm aux}$,
\begin{eqnarray}
 \mathcal{H}_{\textrm{aux},\textrm{eff}} \approx &&
 \nonumber \\
\frac{1}{2} \! \!
\sum_{\mu, \langle \alpha \rangle_\mu} 
&&g^{\mu}_\alpha (t) g^{\mu}_\beta (t)
\Big( 
\frac{1}{\omega_{\rm aux} -\omega_{\alpha}} \!
+\nonumber \\
&&
\!
\frac{1}{\omega_{\rm aux} -\omega_{\beta}}  
\Big)
c_\alpha^\dagger c_\beta 
e^{- i \Delta_{\alpha,\beta} t}  + \text{H.c.} , \label{eq:eff.couplings2}
\end{eqnarray}
where $\Delta_{\alpha,\beta}=\omega_\alpha-\omega_\beta$.
The last expression allows us to control couplings between $\alpha$ and $\beta$ cavities, as long as $\Delta_{\alpha,\beta}$ take different values. Diagonal terms in Eq. \eqref{eq:eff.couplings2} will lead to energy shifts that can be re-absorbed into the definition of the main cavity frequencies.

To simplify our analysis, in view of the limit in Eq. \eqref{eq:limits}, we assume that the variations in cavity frequencies between the denominators can be neglected, such that 
$\omega_\alpha - \omega_{\rm aux} \approx \omega_\beta - \omega_{\rm aux} \approx 
\bar{\omega} - \omega_{\rm aux}$. We will show below that this approximation can be relaxed. 
We will also consider that the couplings between main and auxiliary cavities are equal, $g_\alpha^\mu(t)=g^\mu(t)$, with a multi-tone time-dependence of the form:
\begin{equation}
g^\mu(t)=\sqrt{\omega_{\rm aux} - \bar{\omega}} \sum_{i} A_i \cos(\Omega_i t)\,,
\label{eq:tones}
\end{equation}
where $i=1,2,\dots$ denotes the number of tones with frequency $\Omega_i$ and amplitude $A_i$. Using that form of $g^\mu(t)$, the time-dependent coupling contribution of $\mathcal{H}_{\textrm{aux},\textrm{eff}}$ reads:
\begin{align}
    \frac{(g^\mu(t))^2}{\omega_\mathrm{aux}-\bar{\omega}}&=\sum_{i,j} A_i A_j \cos(\Omega_i t)\cos(\Omega_j t)=\nonumber \\
    &=\sum_{i,j} \frac{A_i A_j}{4}\left(e^{i(\Omega_i+\Omega_j) t}+e^{i(\Omega_i-\Omega_j) t}+\mathrm{H.c.}\right)\,,
\end{align}
Only the terms that satisfy: $\Omega_i\pm\Omega_j=(-) \Delta_{\alpha\beta}$ will be resonant in $\mathcal{H}_{\textrm{aux},\textrm{eff}}$, while the rest will average out to zero in a limit that we describe below. Defining an energy distribution of $\omega_\alpha$ like the one depicted in Fig.~\ref{circuit}(b), with intra-cell detuning $\delta$, and different intra-cell detuning $\delta_i$, the third-neighbour hoppings, e.g., at the $\mu=\mathrm{I}$-resonator, have energy detunings:
\begin{align}
    \Delta_{41}&=2\delta+\delta_1\,,\\
    \Delta_{52}&=\delta_1+\delta_2+\delta\,,\\
    \Delta_{63}&=2\delta+\delta_2\,,
\end{align}
the second-neighbour hoppings:
\begin{align}
    \Delta_{31}=\Delta_{42}&=\delta+\delta_1\,,\label{eq:second}\\
    \Delta_{53}= \Delta_{64}&=\delta+\delta_2\,,
\end{align}
whereas the first-neighbour hoppings:
\begin{align}
    \Delta_{21}=\Delta_{43}=\Delta_{65}&=\delta\,,\\
    \Delta_{32}=\delta_1\,,\, \Delta_{54}&=\delta_2\label{eq:first}\,.
\end{align}

From all these processes, we need to find a tone structure $(\Omega_i,A_i)$ that make resonant only the desired first and  third-order processes, while keeping off-resonant the rest. We choose first three set of tones to activate the third-order tunneling that satisfy:
\begin{align}
    \Omega_1+\Omega_2&=2\delta+\delta_1= \Delta_{41}\,,\label{eq:first_resonance_third}\\
    \Omega_1+\Omega_3&=\delta+\delta_1+\delta_2= \Delta_{52}\,,\\
    \Omega_2+\Omega_3&=2\delta+\delta_2= \Delta_{63}\,\label{eq:last_resonance_third}.
\end{align}
and $A_1 A_2=J_3=A_2 A_3$ and $A_1 A_3=J_3^\prime$, that can be obtained  with $A_1=A_3=\sqrt{J_3^\prime}$ and $A_2=J_3/\sqrt{J_3^\prime}$. At the even auxiliary resonators $\mu=\text{II},\text{IV},\dots$ the role of $J_3$ and $J_3^\prime$ should be reversed so that third-neighbour hoppings have the right alternating structure, $J_3,J_3^\prime,\dots$. 

We also need to add extra multi-tone drivings that activate first-neighbor couplings. The following choice of frequencies can do the job:
\begin{align}
\Omega_4+\Omega_5&=\delta\,, \label{eq:first_resonance_nearest}\\
\Omega_4+\Omega_6&=\delta_1\, , \\
\Omega_5+\Omega_6&=\delta_2\,\label{eq:last_resonance_nearest} ,    
\end{align}
with first-neighbor couplings given by $A_4 A_5 = J_1$, $A_4 A_6 = J^\prime_1 = A_5 A_6$, which can be implemented with the driving amplitudes $A_4 =  A_5 = \sqrt{J_1}$, $A_6 = J^\prime_1 / J_1$. In the Appendix~(\ref{sec:more_experimental_implementation}), we check that the tone frequencies are only resonant with the desired extended SSH couplings, and any undesired hopping is activated.

Under the assumption that all the off-resonant time-dependent terms in $\mathcal{H}_{\textrm{aux},\textrm{eff}}$ oscillate at a much faster scale than their amplitudes, $A_i A_j$, $\mathcal{H}_{\textrm{aux},\textrm{eff}}$ approximates the extended SSH Hamiltonian discussed along this manuscript, $\mathcal{H}_{\textrm{aux},\textrm{eff}}\approx H_B$. Let us finally note that one can relax one of the assumptions made, $\omega_\alpha+\omega_\beta\ll \omega_\mathrm{aux}$, and correct the different values connecting the $\alpha,\beta$ cavities through the amplitudes of the tones, $A_i$.

\section{Conclusion \& Outlook~\label{sec:conclu}}

Summing up, we study the quantum optical consequences of letting quantum emitters interact with a one-dimensional topological photonic bath with phases characterized by a large winding numbers ($\mathcal{W}>1$). When the emitters are coupled to the bulk modes, i.e., at the center of the chain, we show the emergence of qubit-photon bound states with qualitatively different features from the standard SSH model, e.g., with different spatial periodicities, and provide a way of tuning their shape through the use of giant atoms. Besides, we unravel how the photonic component of these bound-states can be understood from the hybrization of topological edge states, and thus inherit their protection to disorder, as we numerically benchmark. Then, we show that by coupling the emitters to the borders of the chain, they can efficiently interact with the topological edge states appearing in the phases with $\mathcal{W}\neq 0$, dominating the spontaneous decay dynamics of single emitters. Interestingly, we find that in the giant atom case, its dynamics becomes more sensitive to the different phases of the bath $|\mathcal{W}|$, since it couples efficiently to all the topological edge states irrespective of the phase, something not possible with local couplings. Finally, we propose a circuit QED architecture to implement these topological light-matter interfaces using Floquet-modulated couplings.

\bibliographystyle{apsrev4-1}
\bibliography{references}

\begin{acknowledgements}
  C. Vega, D. Porras and A. González-Tudela acknowledge   support   from   CSIC Research   Platform   on   Quantum   Technologies   PTI-001  and  from  Spanish  project  PGC2018-094792-B-100(MCIU/AEI/FEDER, EU). M. Bello acknowledges support from the ERC Advanced Grant QUENOCOBA (GA No. 742102). 
\end{acknowledgements}

\appendix
\renewcommand\thefigure{\thesection.\arabic{figure}}    
\setcounter{figure}{0}    

\section{Dynamics of a single emitter in the band regime~\label{sec:emitter_band_dynamics}}

\begin{figure}[tb]
  \centering \includegraphics[width=7.5cm]{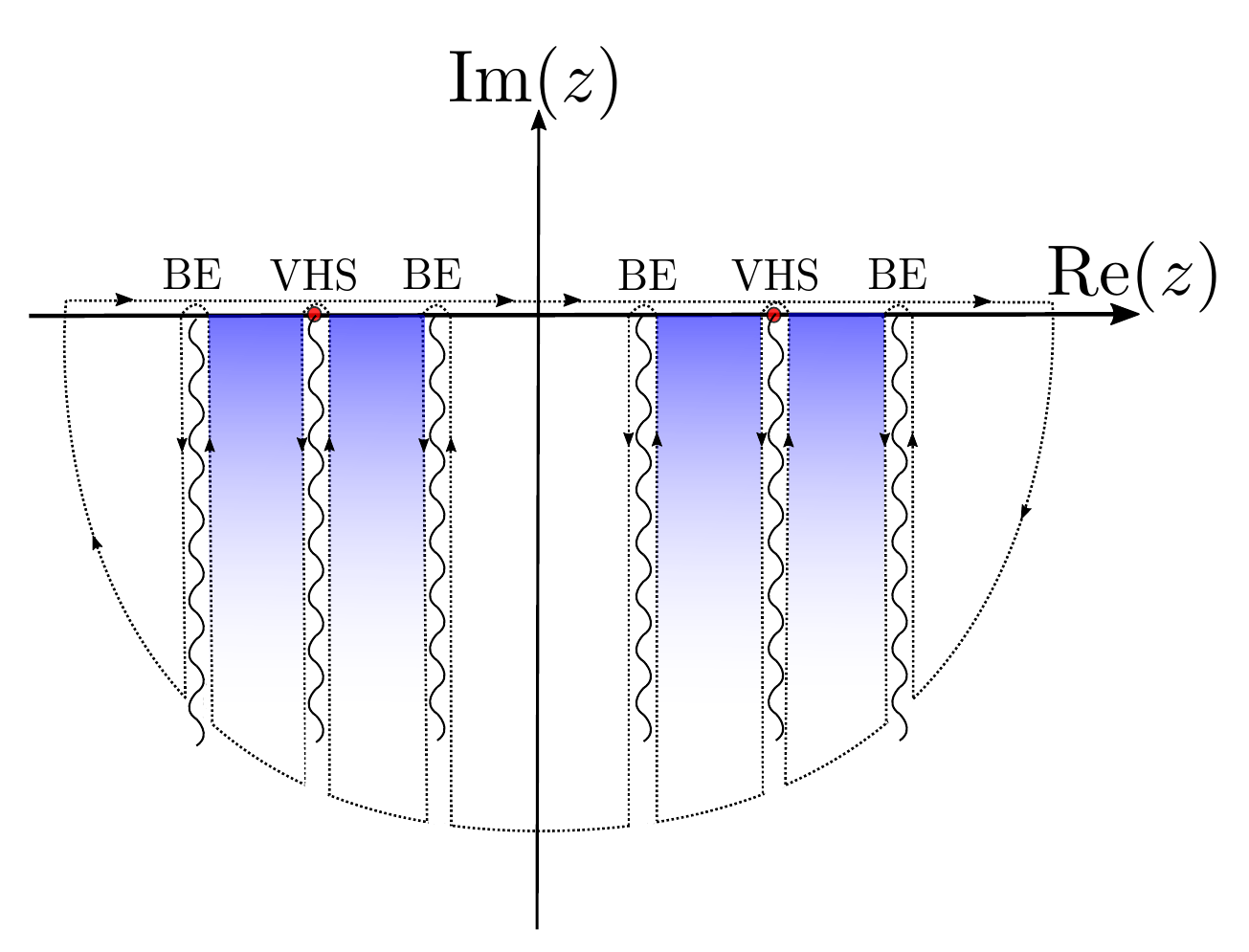}
  \caption{Representation of the integration contour. Blue-shaded regions correspond to the band regime. The Green function has branch cuts at the band edges (BE) as in the SSH model, but it displays extra branch cuts at the band regime due to the emergence of van Hove singularities (VHS) when long-range hoppings are large enough.}
  \label{branch_cuts}
\end{figure}

Generally, if an excited emitter is coupled in resonance to a bulk band of a photonic lattice in its vacuum state, the emitter will radiate away its excitation in a Markovian fashion. However, there are situations in which this decay is non-Markovian, displaying a backflow of the excitation from the lattice to the emitter. In particular, in the manuscript body we discussed that the extended SSH band structure exhibits a van Hove singularity within the band energies if long-range hopping amplitudes are large enough. If we consider an emitter frequency at the van Hove singularity, the density of states diverges and, therefore, its dynamics is expected to deviate significantly from a Markovian one~\cite{Gonzalez-Tudela2017a,Gonzalez-Tudela2017b}.\\

To check it, we compute the emitter dynamics within the resolvent operator formalism. This method allows to describe the emitter evolution in terms of the poles of the Green function associated to the single emitter self-energy $\Sigma_e(z)$, i.e, $G_e(z)=(z-\Delta-\Sigma_e(z))^{-1}$. Although long-range hoppings significantly complicate an analytical derivation of $\Sigma_e(z)$, we can take a semi-analytical approach to estimate the long-term evolution of $|C_e(t)|^2$. The way to do it consists in writing the excited-state probability amplitude, $C_e(t)$, as the inverse Laplace transform of the single-emitter Green function $G_e(z)$:
\begin{equation}
    C_e(t)=-\frac{1}{2\pi i}\int_{-\infty}^{\infty}dz\frac{e^{-izt}}{z+i 0^+-\Delta-\Sigma_e(z+i 0^+)}\;.
\end{equation}

This integration can be performed using complex analysis techniques by choosing the contour depicted in Fig.~\ref{branch_cuts} and applying the Residue Theorem. Note, we have to perform several detours in the contour to avoid the non-analytical regions of the self-energy. Using that method, the $C_e(t)$ can be written as a sum of several contributions:
\begin{equation}
    C_e(t)=\sum_{\text{BS}}R(z_\text{BS})e^{-iz_\text{BS}t}+\sum_{\text{UP}}R(z_\text{UP})e^{-iz_\text{UP}t}+\sum_{\text{BC}}C_\text{BC}(t)\;.
\end{equation}

The first two terms correspond to the contribution of the real and unstable poles of the Green Function. The real ones appear at the qubit-photon bound states (BS) energies along the real axis, $z_\text{BS}=E_\text{BS}$, whereas the unstable poles (UP) are complex and lead to exponential decay dynamics. The function $R(z)$ denotes the residues computed at each pole $z$. Apart from these two terms, there are additional non-exponential decay terms associated to the branch-cut (BC) detours. Differently from other one-dimensional models where these BC only appear at the band-edges, in the extended SSH model with large long-range hoppings there exist also in-band BCs due to the existence of a Van-Hove singularity, as represented in Fig.~\ref{branch_cuts}. If we energetically tune the emitter near this Van-Hove value, $x_\star$, its dynamics will be dominated by this term. Under this assumption, the long-term evolution of the emitter population can then be approximated by:
\begin{equation}
    \lim_{t\to\infty} |C_e(t)|^2 \approx \left|\int_0^\infty dy\; F_\star(y)e^{-yt}e^{-ix_\star t}\right|^2\,,
    \label{branch_cut_dynamics}
\end{equation}
where $x_\star$ is the van Hove singularity energy in the upper band and the function $F_\star(y)$ is given by
\begin{equation}
    F_\star(y) = \frac{2\Sigma_e(x_\star-iy)}{(x_\star-iy-\Delta)^2-\Sigma_e^2(x_\star-iy)}\;.
\end{equation}

Since $F_\star(y)$ is exponentially suppressed in the integrand of Equation~(\ref{branch_cut_dynamics}) for large $t$, the long-time dynamics of $C_e(t)$ is dominated in this case by the behaviour of $F_\star(y)$ near $y=0$. With a numerical fitting (not shown) we find $F_\star(y)\sim y^{\alpha}$, with $\alpha\approx 1.22$, with which we can approximate long time-decay dynamics using
\begin{equation}
    \lim_{t\to\infty}|C_e(t)|^2 \propto \left|\int_0^\infty dy\; y^{\alpha}e^{-yt}e^{-ix_\star}\right|^2 \propto \frac{1}{t^{2(1+\alpha)}}\;,
\end{equation}

\begin{figure}[tb]
\includegraphics[width=8.5cm]{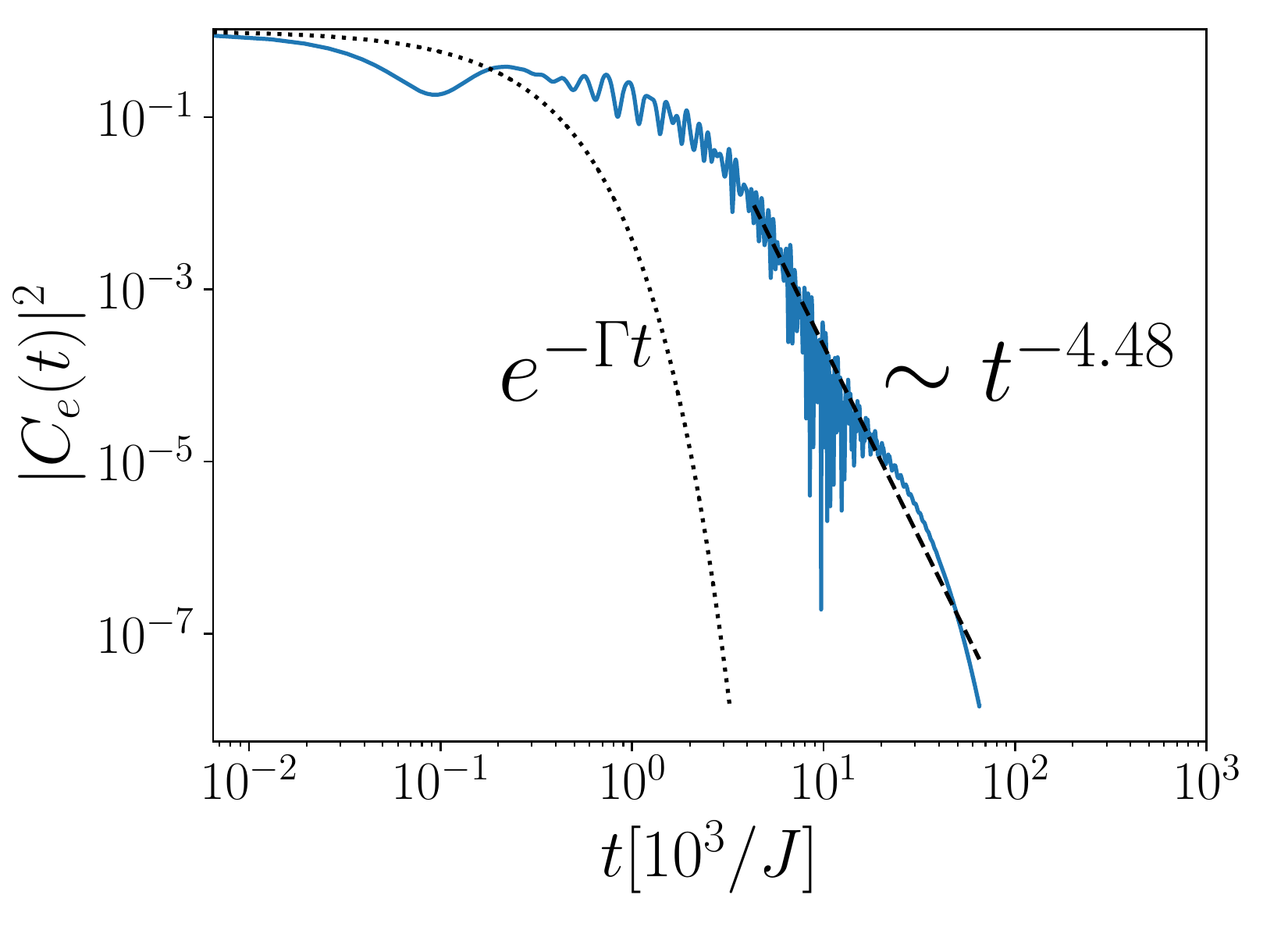}
\caption{Dynamics of a single emitter tuned at the van Hove singularity $\Delta_\text{VHS}\approx 5.03J$ in an extended SSH model configuration with hoppings of $(J_1,J_1^\prime,J_3,J_3^\prime)=(1,1,2,4)J$ and a bath-emitter coupling constant $g=0.1J$ (solid blue line). The dotted line depicts the Markovian approximation $e^{-\Gamma t}$, with $\Gamma$ calculated from the imaginary part of the self-energy defined in Eq.~\ref{self_energy_eq2}. The dashed line is a guided to the eye depicted the expected algebraic decay with $\sim t^{-4.48}$.}
\label{van_Hove_dynamics}
\end{figure}

Thus, using the numerically obtained $\alpha\approx 1.22$, one should expect an algebraic decay rate $t^{-\beta}$ of $\beta\approx2(1+1.22)=4.44$, which is what we find when simulating numerically the full emitter+bath dynamics using the time-dependent Schrödinger equation, as shown in Fig.~\ref{van_Hove_dynamics}.

\section{On the shape of qubit-photon bound states~\label{sec:BS_shape}}

To study in detail the bound-state shape at $E_\mathrm{BS} = 0$, we can particularize Eqs.~\eqref{eq:bsa} and \eqref{eq:bsb} to that energy, which yields:
\begin{equation}
    C^A_{j,a} = 0 \,, \quad
    C^A_{j,b} \propto \frac{1}{2\pi} \int dk\, \frac{e^{ikj}}{f(e^{ik})} \,,
\end{equation}
where $f(e^{ik}) = J^\prime_3 e^{ik} + J^\prime_1 + J_1 e^{-ik} + J_3 e^{-2ik}$. Doing a change of variable $y = \exp[\text{sign}(j)ik]$, we can transform the integral giving the amplitudes $C^A_{j,b}$ into an integral along the unit circumference (anti-clockwise) in the complex plane of a rational function of $y$, which can be integrated by residues, yielding
\begin{equation}
    C^A_{j,b}\propto \sum_{|y_\alpha|<1} y^{|j| + (s + 1)/2} \text{Res}(y_\alpha)\,.
    \label{eq:bsanalitical}
\end{equation}
Here, $s\equiv \text{sign}(j)$, and $y_\alpha$ are the roots of the third-degree polynomial $y^{(s+3)/2} f(y^s)$, i.e., they are the roots of $p(y)\equiv y^2 f(y)$ if $j\geq 0$, or the roots of the reciprocal polynomial $p^*(y)\equiv y^3p(y^{-1})=yf(y^{-1})$ if $j<0$. The residues, $\text{Res}(y_\alpha)$, correspond to the residues of $1/p(y)$ or $1/p^*r(y)$ accordingly. Note that $p^*$ is the same as $p$ interchanging $J_n\leftrightarrow J^\prime_n$. Also note that the roots of $p$ are the inverses of those of $p*$.

We are now in a good position to discuss the different features of the bound states. First, in the thermodynamic limit, the bound state only has weight in one of the bath's sublattices. It is localized around the emitter in a way dictated by the roots of $p$. For example, let us assume that the roots are all different, if all of them (none of them) lie within the unit circle, the bound state will display a multi-exponential decay just on the right (left) side of the emitter, and it will vanish completely on the opposite side. This happens in the phases with $\mathcal{W} = 2, -1$. On the other hand, If one (two) roots lie within the unit circle, it will decay exponentially on the right (left) and multi-exponentially on the left (right) of the emitter. This happens in the phases with $\mathcal{W} = 0, 1$.  Remarkably, there is an interesting relationship between the number of poles within the unit circle and the winding number given by: $\mathcal{W}=2-\#(\text{poles of $p$ within the unit circle})$.

\begin{figure}
    \centering
    \includegraphics[width=\linewidth]{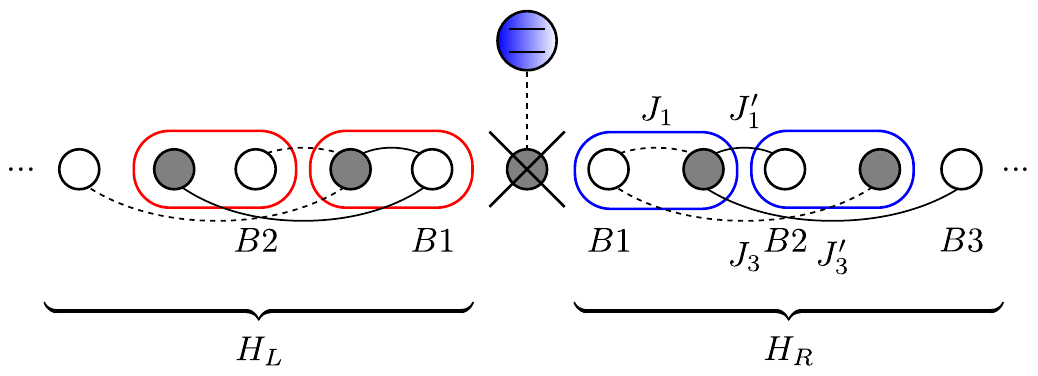}
    \caption{Schematic drawing showing the two semi-infinite chains whose edge modes participate in the vacancy mode (photonic part of the bound-state wavefunction). Note how the hoppings in the left/right chains are reversed, $J_n\leftrightarrow J'_n$}
    \label{figA:semichains}
\end{figure}

Now, we will show the connection of the photonic component of the bound-state with the edge states of the extended SSH model that are obtained when introducing a vacancy at the emitters position, as explained first in Ref.~\cite[Supplementary Material]{Leonforte2020b}. For this, we have to consider the two semi-infinite chains that result when we split the bath at the emitter position, each one described by the Hamiltonians $H_{L/R}$ (see Fig.~\ref{figA:semichains}). We can use the ansatz $\ket{\psi_{\mathrm{ES},\alpha}} = \sum_n \xi_\alpha^n b^\dagger_n \ket{\mathrm{vac}}$ to find the edge states of these chains. The secular equations $H_{L/R}\ket{\psi_\text{ES}} = 0$ impose the conditions $J_3\xi_\alpha^{n-2} + J_1\xi_\alpha^{n-1} + J'_1\xi_\alpha^n + J'_3\xi_\alpha^{n+1} = 0$ for the right chain, or the same equation changing $J_n\leftrightarrow J^\prime_n$ in the left chain. In other words, $\xi_\alpha$ is a pole of $p$ or $p^*$. Furthermore, normalization of the edge state requires $|\xi_\alpha| < 1$. Comparing this with the expression in Eq.~\eqref{eq:bsanalitical}, it is clear now that the photonic component of $\ket{\psi_\text{BS}}$ is an exact superposition of these edge states. Let us also finally note that in small finite systems with open boundary conditions the bound states can hybridize significantly with other edge states of the chain and delocalize from the emitter position. Besides, they can also have a non-negligible contribution from the bulk modes.

\begin{figure}[htb]
\includegraphics[width=9cm]{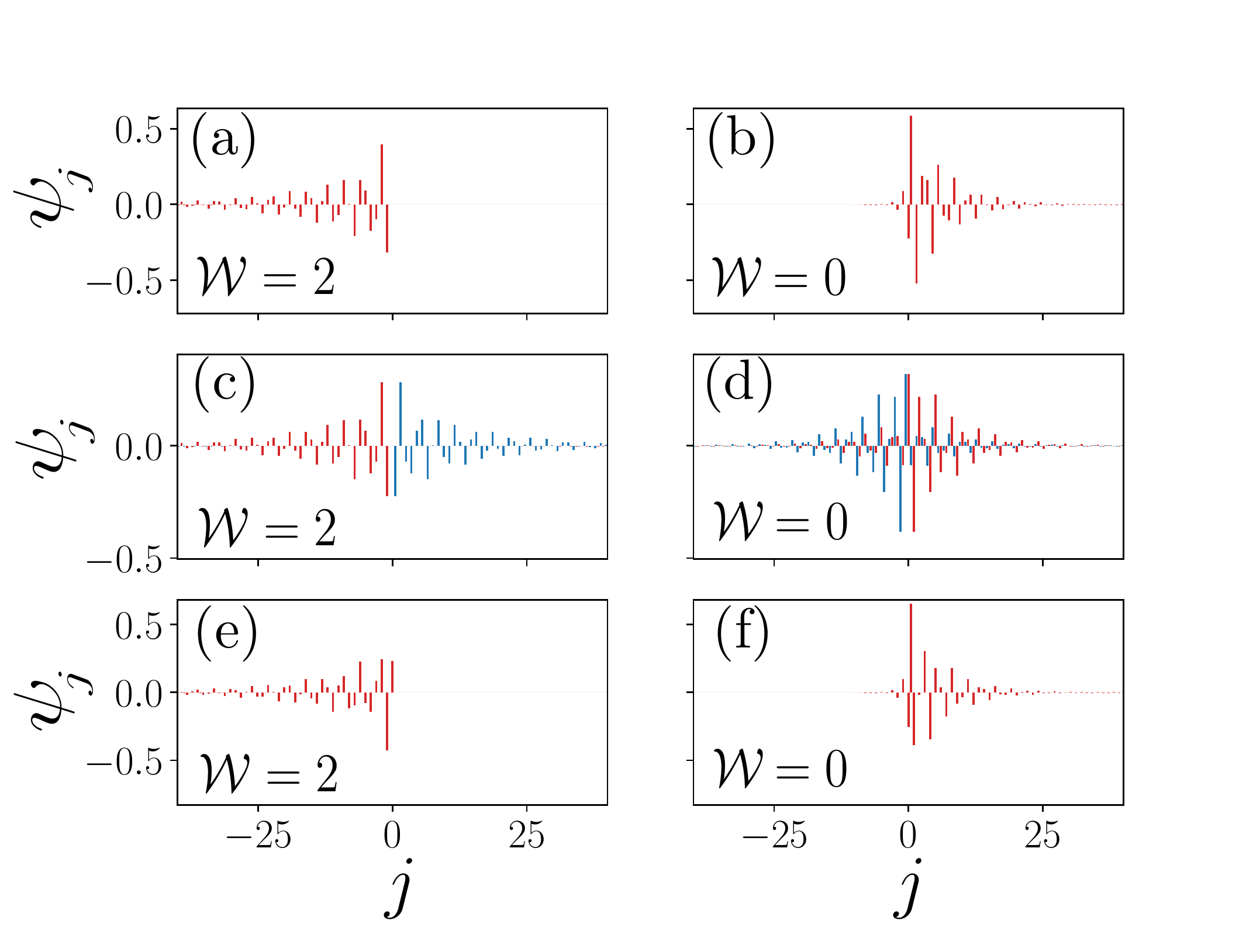}
\caption{Representation of vacancy-like dressed states (VDS) for a single vacancy placed at site $A$ in unit cell $j=0$ [(a) and (b)], two vacancies at sites $A$ and $B$ in unit cell $j=0$ [(c) and (d)], and two vacancies at sites $A$ in unit cells $j=0$ and $j=1$ [(e) and (f)]. The bath hamiltonian parameters are $(J_3^\prime,J_3)=(0.5J, 0.8J)$ in $\mathcal{W}=2$ and $(J_3^\prime,J_3)=(0.2661J, 0.5J)$ in $\mathcal{W}=0$. Plots (a) and (b) resemble the middle band-gap bound state wavefunctions for an emitter with a local coupling, depicted in Fig.~\ref{shape_bound_states_diff_detunings}; whereas panels (c) to (e) resemble the bound state structure when coupling a giant atom to the lattice bulk, represented in Fig.~\ref{giant_atoms}}
\label{vacancy}
\end{figure}

Finally, let us also mention that the vacancy-like dressed state (VDS) identification of Ref.~\cite{Leonforte2020b} works both for the local and non-local couplings situation that we have discussed along the manuscript. In fact, we also show in Fig.~\ref{vacancy} that the zero-energy modes of the vacancy-like Hamiltonian $\mathcal{H}_{B,v}$ have the same spatial shape of $\ket{\psi_\mathrm{ph}}$ plotted in the main manuscript for both local and non-local coupling cases in all topological phases. From this, we can numerically evidence that indeed the vacancy-like modes of $\mathcal{H}_{B,v}$ result from the hybridization of the topological edge states of the uncoupled chains, $H_R+H_L$, through the long-range hoppings.

\section{Lack of topological protection in outer band-gaps and localization effects~\label{sec:disorder_driven}}

\begin{figure}[tb!]
\includegraphics[width=\columnwidth]{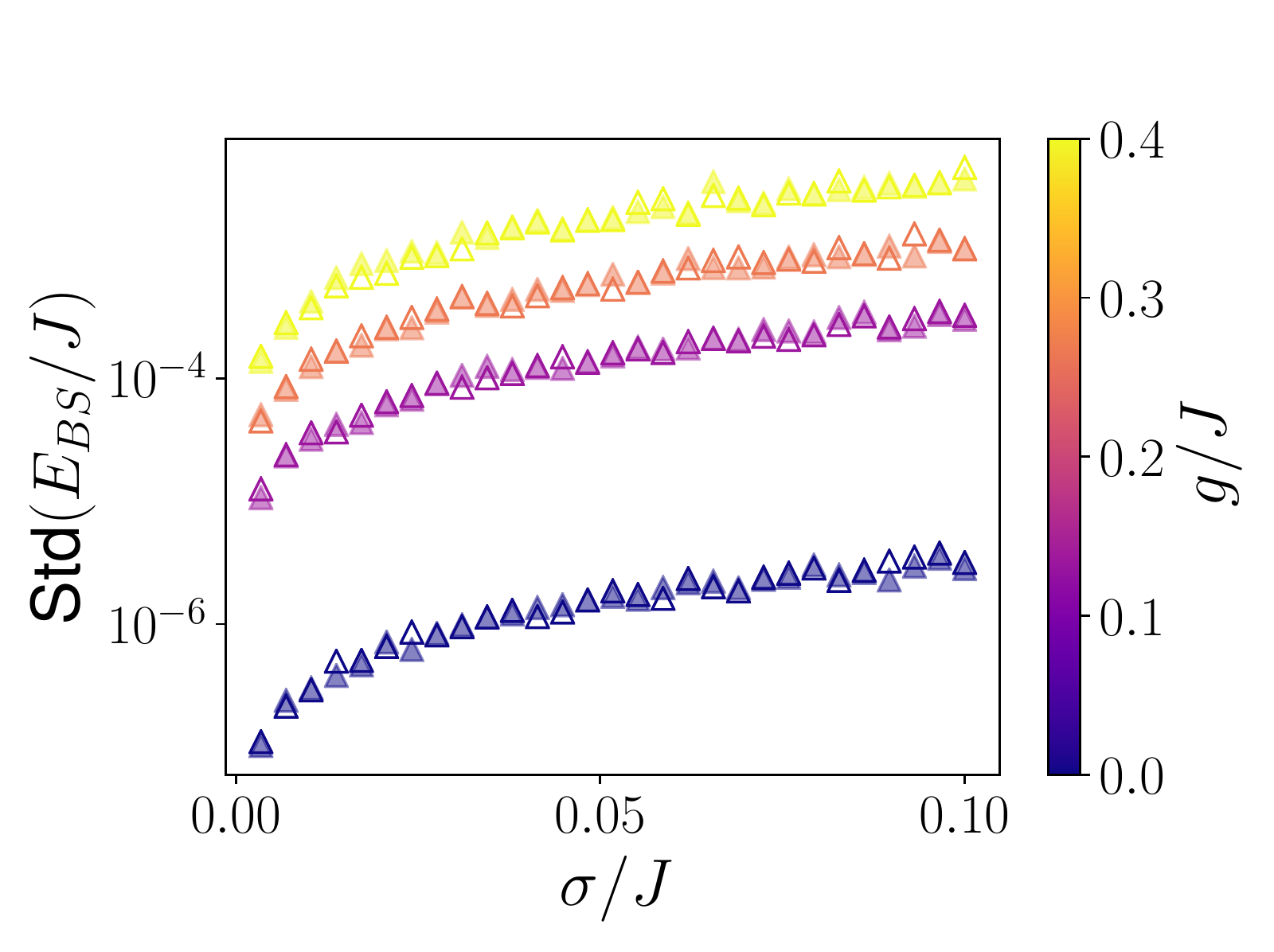}
\caption{Standard deviation of lower band-gap qubit-photon bound state energies $E_\text{BS}$ as gaussian disorder strength $\sigma$ is increased, for different values of the coupling constant $g$ (in different colors) for lattice sizes of $N=300$ (empty triangles) and $N=600$ (full triangles) sites. The parameters of both models have been set as in Fig.~\ref{bandgap_protection}. In both figures, each point corresponds to the standard deviation in a sample of $200$ realizations of chirality-preserving disorder with fixed $\sigma$.}
\label{lower_variances}
\end{figure}

In the main text, we conclude that the outer band-gap bound states in the extended SSH model lack topological protection, from the results depicted in Fig~(\ref{disordered_BS_energy}). Here, we will show this more explicitly by exploring the variance of the $E_\text{BS}$ distribution for different bath sizes, as we did for the middle band-gap bound states in the main text. In that case, we use the decrease of $\text{Std}(E_\text{BS}(\sigma))$ as system size increases as a strong indication of the topological protection of these bound states. In Fig.~\ref{lower_variances} we make a similar analysis for the lower band-gap bound states by plotting $\text{Std}(E_\text{BS}(\sigma))$ as disorder increases for different system sizes (empty/filled markers) and several coupling strengths (in different colors). There, we observe that these variances remain fixed regardless of the lattice size. This resembles the behaviour of middle band-gap bound states in the topologically trivial staggered-energy model. Thus, we conclude that outer band-gap bound states are indeed not topologically protected in spite of the topological nature of the bath.

Another interesting observable to monitor as disorder increases is the localization length of the qubit-photon bound states. For characterizing this property we can use the Inverse Participation Ratio (IPR), which is defined for a state written in real space $|\psi\rangle=\sum_j c_j|j\rangle$ as
\begin{equation}
    \text{IPR}(\psi) = \frac{1}{\sum_j |c_j|^4}
\end{equation}

\begin{figure}[tb!]
\includegraphics[width=\columnwidth]{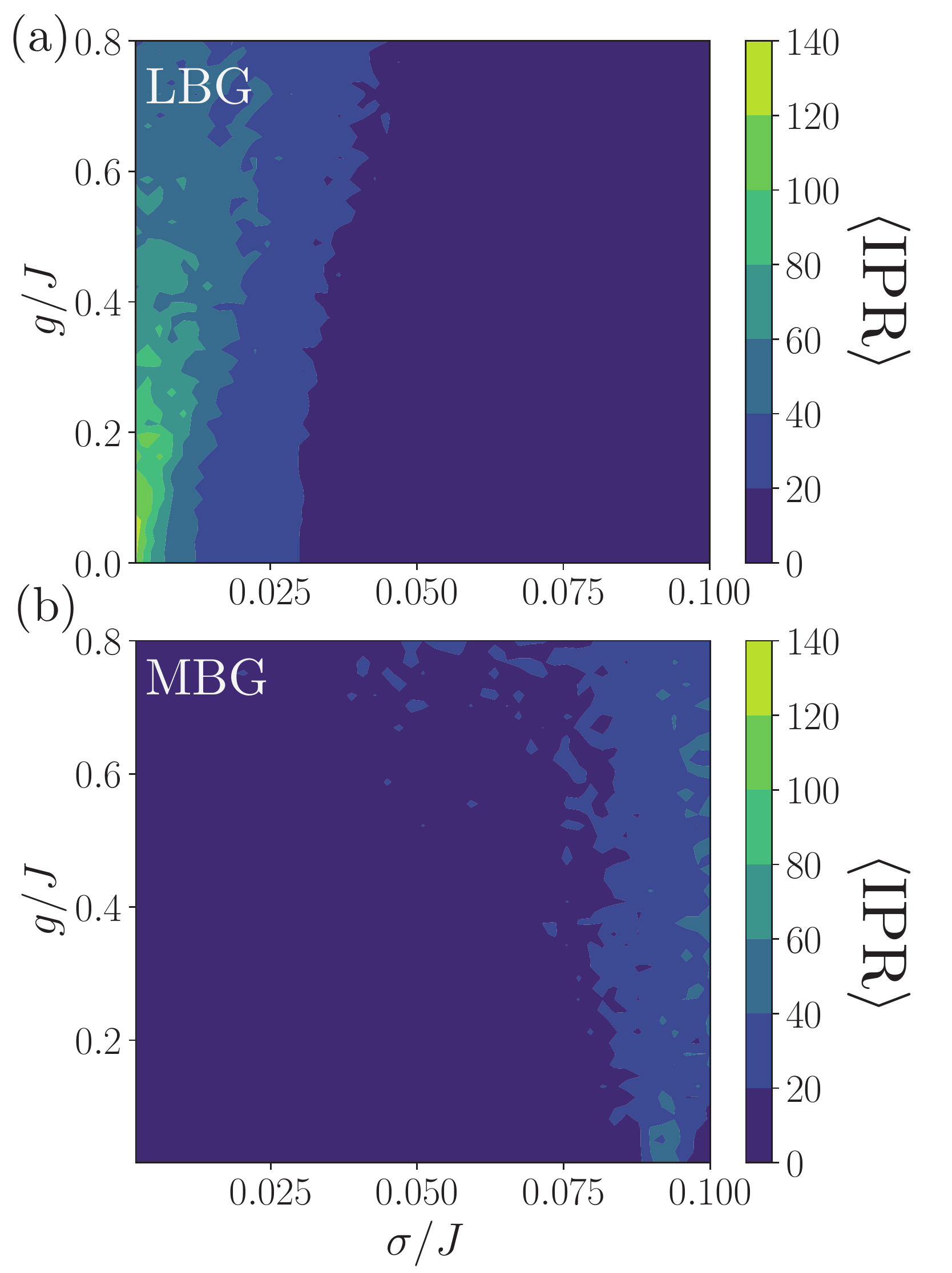}
\caption{Mean inverse participation ratio (IPR) over $50$ disorder samples for qubit-photon bound states in a) the lower ($\Delta/J=-3.3$) and b) middle ($\Delta/J=0$) band-gaps of the extended SSH model, with varying coupling constant and disorder strength. The model parameters are set to $(J_3^\prime,J_3)=(0.5,0.8)$ and the lattice size is of $N=200$ sites.}
\label{IPR}
\end{figure}

The larger the IPR, the less localized is the bound-state (and viceversa). In Fig.~\ref{IPR} we represent the IPR for qubit-photon bound states at the lower and middle band-gaps under chirality-preserving disorder. In particular, we plot the IPR as a function of the disorder strength (horizontal axis) and coupling strength (vertical axis) using a color scale (see legend). We can observe that lower band-gap bound states tend to localize as disorder increases, behaviour that is shared with bulk waves, which is a signature of Anderson localization. On the other hand, we observe that middle band-gap bound states display a more robust IPR for weak disorder, and begins to delocalize for disorder strengths close to $E_g/2$ in the weak coupling regime ($g\ll J$).

\section{Effective finite-bath dynamics~\label{sec:effective_model}}

\begin{figure}[tb!]
\includegraphics[width=9cm]{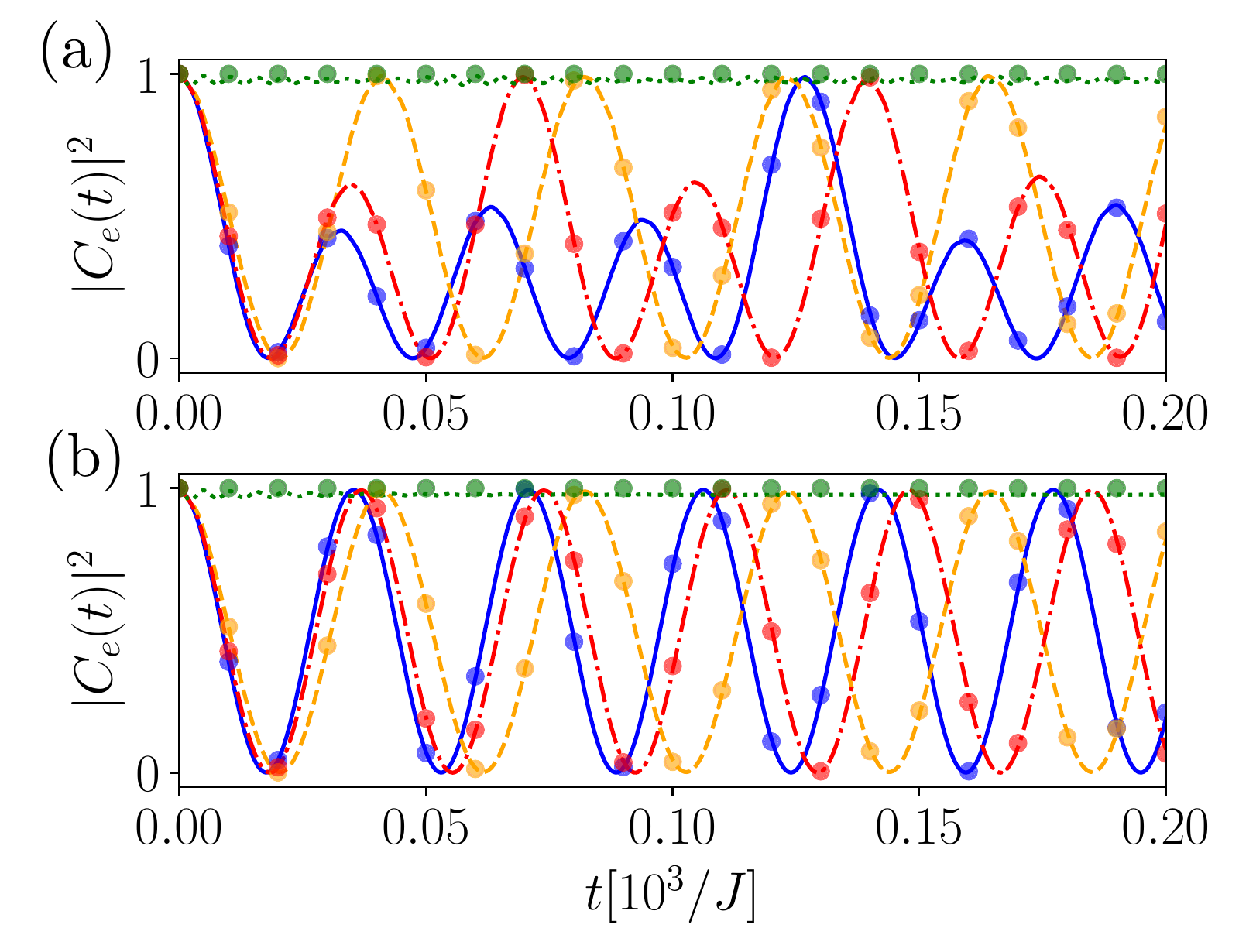}
\caption{Dynamics of a giant atom coupled to the left edge of an extended SSH lattice of (a) $N=20$ and (b) $N=120$ sites. Lines represent the dynamics as given by the full light-matter interaction hamiltonian, while markers represent the predictions of the effective model we have described. All parameters and color codes are the same as in Fig.~\ref{edge_dynamics_1} }
\label{effective_fourier_dynamics}
\end{figure}

In the body of this manuscript we have discussed the dynamics of emitters coupled at the ends of an extended SSH chain using the full Hamiltonian of the model. However, when $\Delta$ lies in the band-gap, we expect that edge states are the only bath modes that will contribute to the dynamics of the emitters. This allows us to formulate an effective model that approximates the emitters dynamics projecting into the edge state subspace generated by $|e\rangle|\text{vac}\rangle$ and $\left\lbrace|g\rangle|\text{ES}_i\rangle\right\rbrace$, where $|\text{ES}_i\rangle$ are now the edge state of the complete bath Hamiltonian $\mathcal{H}_B$ (unlike in Section~\ref{sec:BS_shape}), and where the index $i$ denotes the multiplicity of the edge states. Using this projection, the generic form of the effective Hamiltonian reads:
\begin{equation}
\mathcal{H}_\text{eff}=\Delta|e\rangle\langle e| + \sum_{i}\varepsilon_i|\text{ES}_i\rangle\langle\text{ES}_i| + \Tilde{g}_i |\text{ES}_i\rangle\langle e|+\text{H.c}.    
\end{equation}

The term $\varepsilon_i$ represents the energy of the edge state $|\text{ES}_i\rangle$, while $\Tilde{g}_i$ denotes the effective coupling constant between the emitter and the edge state $\Tilde{g}_i=\langle \text{ES}_i|\mathcal{H}_\text{int}|e\rangle$. In the case of a local emitter-bath coupling, $\Tilde{g} = g|\psi_\text{edge}(j_\text{emitter})|$, being $|\psi_\text{edge}(j_\text{emitter})|$ the spatial support of the edge mode on the site where the emitter is coupled to. If the emitter couples to more that one site, $\Tilde{g}_i$ is the sum of the wavefunction supports in the coupled sites. The number of edge states depends on the absolute value of the winding number, which is the topological invariant characterizing the phase of the bulk hamiltonian. For a winding number $\mathcal{W}=0$ the absence of edge modes leads to a trivial effective model where the emitter can not decay to any channel.

In what follows, we will benchmark the effective model by comparing with the results of Fig.~\ref{edge_dynamics_1} of the main text obtained through the full numerical evolution of the system. This comparison is shown in Fig.~\ref{effective_fourier_dynamics} where we plot the dynamics of a giant atom coupled to the edge in the same conditions described Fig.~\ref{edge_dynamics_1} computed with the full (lines) and effective Hamiltonian (markers). There we observe how indeed the effective model captures the emitter dynamics with a very good agreement, improving as the system size increases. In fact, assuming $\Delta=0$ in the thermodynamic limit we can obtain an analytical approximation of the dynamics for all topological phases $\mathcal{W}$ given by
\begin{equation}
    C_e(t) \sim \cos(\Tilde{g} t)\;, \;\text{with }\Tilde{g} = \sqrt{\sum_i \Tilde{g}_i^2}\;,
    \label{single_emitter_freq}
\end{equation}
where the sum is performed over the edge states. Thus, the larger the number of edge states the emitter couples to, the larger will be the Rabi oscillation.

\begin{figure}[tb]
\includegraphics[width=\linewidth]{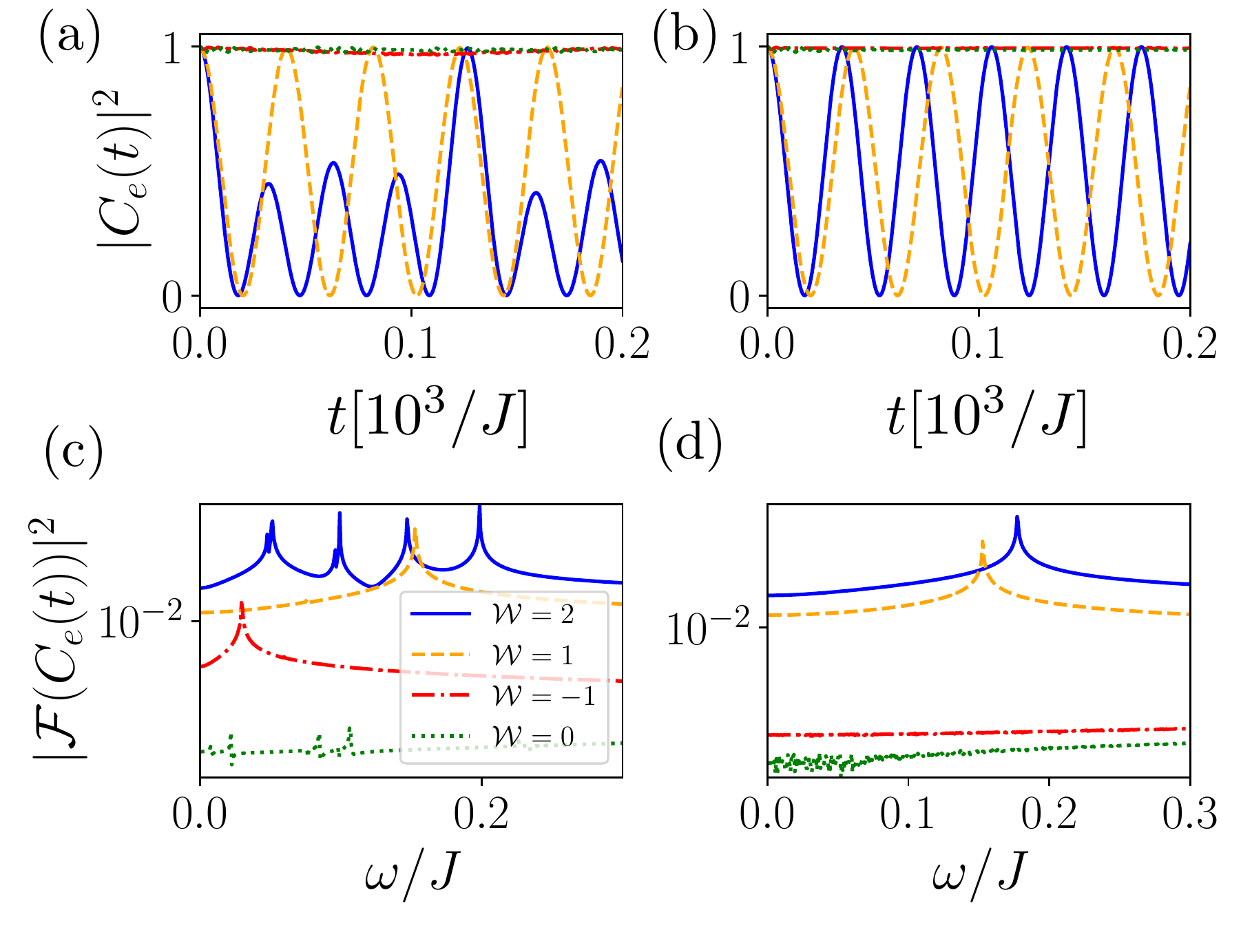}
\caption{Dynamics in time and frequency domains of an emitter with $\Delta/J=0$ locally coupled to the leftmost site of the lattice, for sizes of $N=120$ (a) and c)) and $N=20$ unit cells (b) and d)). All parameters are set as in Fig.~\ref{edge_dynamics_1} of the main text, except of the coupling locality.}
\label{dynamics_local_coupling}
\end{figure}

As a final remark beyond the effective model, let us here justify the suitability of giant atoms for detecting topological phases as compared to emitters with local couplings. For that, in  Fig.~\ref{dynamics_local_coupling}, we compute excited-state dynamics (and its Fourier transform) of an emitter locally coupled to the sublattice A site in the leftmost cell of the lattice using the same lattice parameters than in Fig.~\ref{edge_dynamics_1}. We can observe that the dynamical features in the $\mathcal{W}=-1$ does not exhibit any interaction with the bath, resembling a topologically trivial scenario for large lattice sizes. The reason for this behaviour lies in the chiral symmetry of the bath, that leads to single-sublattice support of the edge modes, and effectively decouples the emitter of the topological edge states. We can notice however that for small lattices, the Fourier transform captures a peak from a very weak oscillation which is not visible in the emitter population dynamics. We can understand this phenomenon using the effective model. In a topological state with $|\mathcal{W}|=1$ with two edge states with energies $\pm\varepsilon$ and effective coupling constants of strength $|\tilde{g}|$, the effective model predicts an evolution given by:
\begin{equation}
    C_e(t)\approx \frac{\varepsilon^2}{\varepsilon^2+2\tilde{g}^2}+\frac{2\tilde{g}^2}{\varepsilon^2+2\tilde{g}^2}\cos\left(\sqrt{\varepsilon^2+2\tilde{g}^2}t\right)
\end{equation}

In the $\mathcal{W}=-1$ with local coupling, we have $\tilde{g}=0$. However, if the lattice size is small (comparable with the edge state localization length) the edge state may have a small but non-vanishing support in the site the emitter is coupled to. In particular, if we have $\tilde{g}\ll \varepsilon$, the effective model predicts a weak oscillation of amplitude $2\tilde{g}^2/(\varepsilon^2+2\tilde{g}^2)\ll 1$.

\section{Absence of resonances in the experimental implementation~\label{sec:more_experimental_implementation}}
In the manuscript, we proposed an experimental implementation of the extended SSH model through cavity-mediated interactions. In particular, we used a time-dependent coupling described in Eq.~(\ref{eq:tones}). In this section, we explicitly show that such set-up does not lead to undesired resonances i.e. that only extended SSH hoppings are activated. We have used six tones, $\Omega_{1,2,3}$ and $\Omega_{4,5,6}$ to modulate third and nearest-neighbour hoppings respectively. Once the tone frequencies are set, we need to check all terms of the form $\Omega_i\pm\Omega_j$ aiming to verify that any of these terms is equal to $(-)\Delta_{\alpha\beta}$, which would yield to an undesired hopping term in the simulated hamiltonian. From Eqs.~(\ref{eq:first_resonance_third})-~(\ref{eq:last_resonance_third}), we can see that tones governing third-neighbour hoppings do not 
\begin{align}
\Omega_2-\Omega_3 &=\delta-\delta_2\\
\Omega_1-\Omega_3 &=\delta_1-\delta_2\\ 
\Omega_1-\Omega_2 &=\delta_1-\delta
\end{align}
Similarly, regarding the $\Omega_{4,5,6}$ tone frequencies, from Eqs.~(\ref{eq:first_resonance_nearest})-~(\ref{eq:last_resonance_nearest}) we can also see that there are no undesired resonances:
\begin{align}
\Omega_4-\Omega_5 &=\delta_1-\delta_2\\
\Omega_5-\Omega_6 &=\delta-\delta_1\\ 
\Omega_6-\Omega_4 &=\delta_2-\delta
\end{align}
Aiming to check if there is any other resonance, we need to compute as well $\Omega_i\pm\Omega_j$ with $i=1,2,3$ and $j=4,5,6$. Regarding the $\Omega_i-\Omega_j$ terms:
\begin{align}
\Omega_1-\Omega_4 &=\frac{1}{2}\left(\delta_1+\delta_2\right)\\
\Omega_1-\Omega_5 &=\frac{1}{2}\left(3\delta_1-\delta_2\right)\\ 
\Omega_1-\Omega_6 &=\frac{1}{2}\left(2\delta+\delta_1-\delta_2\right)\\
\Omega_2-\Omega_4 &=\frac{1}{2}\left(2\delta-\delta_1+\delta_2\right)\\
\Omega_2-\Omega_5 &=\frac{1}{2}\left(2\delta+\delta_1-\delta_2\right)\\ 
\Omega_2-\Omega_6 &=\frac{1}{2}\left(4\delta-\delta_1-\delta_2\right)\\
\Omega_3-\Omega_4 &=\frac{1}{2}\left(3\delta_2-\delta_1\right)\\
\Omega_3-\Omega_5 &=\frac{1}{2}\left(\delta_2+\delta_1\right)\\ 
\Omega_3-\Omega_6 &=\frac{1}{2}\left(2\delta+\delta_2-\delta_1\right)
\end{align}
Finally, regarding the $\Omega_i+\Omega_j$ terms:
\begin{align}
\Omega_1+\Omega_4 &=\frac{1}{2}\left(2\delta+3\delta_1-\delta_2\right)\\
\Omega_1+\Omega_5 &=\frac{1}{2}\left(2\delta+\delta_1+\delta_2\right)\\ 
\Omega_1+\Omega_6 &=\frac{1}{2}\left(3\delta_1+\delta_2\right)\\
\Omega_2+\Omega_4 &=\frac{1}{2}\left(4\delta+\delta_1-\delta_2\right)\\
\Omega_2+\Omega_5 &=\frac{1}{2}\left(4\delta-\delta_1+\delta_2\right)\\ 
\Omega_2+\Omega_6 &=\frac{1}{2}\left(2\delta+\delta_1+\delta_2\right)\\
\Omega_3+\Omega_4 &=\frac{1}{2}\left(2\delta+\delta_1+\delta_2\right)\\
\Omega_3+\Omega_5 &=\frac{1}{2}\left(2\delta+3\delta_2-\delta_1\right)\\ 
\Omega_3+\Omega_6 &=\frac{1}{2}\left(3\delta_2+\delta_1\right)
\end{align}
We observe that any of the $\Omega_i\pm\Omega_j$ fits any $(-)\Delta_{\alpha\beta}$, meaning that only extended SSH couplings are activated.
\end{document}